\documentclass{aa}  

\usepackage{graphicx}
\usepackage{newtxtext}
\usepackage[varvw]{newtxmath}
\usepackage{booktabs}
\usepackage{bm}
\usepackage{nicefrac}
\usepackage{mathtools}
\usepackage[hidelinks]{hyperref}
\hypersetup{
  colorlinks,
  citecolor=blue,
  linkcolor=blue,
  urlcolor=blue}
\setcitestyle{citesep={;}}
\usepackage{bigints}
\usepackage{orcidlink}

\usepackage{amstext} 
\usepackage{array}   
\newcolumntype{L}{>{$}l<{$}} 
\newcolumntype{R}{>{$}r<{$}} 

\newcommand{\dd}{\ensuremath\mathrm{d}}

\newcommand{\iso}[2]{\ensuremath ^{#1}\mathrm{#2}}

\begin{document}

   \title{
Heavy element abundances from a universal primordial distribution}
   \subtitle{}

\titlerunning{Site of heavy element synthesis}

   \author{G. R{\"o}pke
          \inst{1}\orcidlink{0000-0001-9319-5959}
          \and
          D. Blaschke\inst{2,3,4}\orcidlink{0000-0002-8399-5183}
          \and
          F. K. R{\"o}pke\inst{5,6}\orcidlink{0000-0002-4460-0097}
          }

    \institute{Universit{\"a}t Rostock, Institute of Physics,
              Albert-Einstein-Str.~23, D-18057 Rostock\\
              \email{gerd.roepke@uni-rostock.de}
    \and
        Institute of Theoretical Physics, University of Wroclaw, Max Born place 9, 50-204 Wroclaw, Poland
    \and
        Helmholtz-Zentrum Dresden-Rossendorf (HZDR), Bautzener Landstrasse 400,
        01328 Dresden, Germany
    \and
        Center for Advanced Systems Understanding (CASUS), Untermarkt 20, 
        02826 G\"orlitz, Germany
    \and
        Zentrum f{\"u}r Astronomie der Universit{\"a}t Heidelberg, Institut f{\"u}r Theoretische Astrophysik, Philosophenweg 12, 
        69120 Heidelberg, Germany
    \and
        Heidelberger Institut f{\"u}r Theoretische Studien, Schloss-Wolfsbrunnenweg 35, 69118 Heidelberg, Germany
             }

\authorrunning{G.~R{\"o}pke et al.}

 \date{\today}

\abstract
{We present a freeze-out approach to the formation of heavy elements in expanding nuclear matter. Applying concepts used in the description of heavy-ion collisions or ternary fission, we determine the abundances of heavy elements taking into account in-medium effects such as Pauli blocking and the Mott effect, which describes the dissolution of nuclei at high densities of nuclear matter. With this approach, we search for a  universal primordial distribution in an equilibrium state from which the gross structure of the solar abundances of heavy elements freezes out via radioactive decay of the excited states. The universal primordial state is characterized by the Lagrangian parameters of temperature and chemical potentials of neutrons and protons. We show that such a state exists and  determine a temperature of $5.266\, \mathrm{MeV}$, a neutron chemical potential of $940.317\, \mathrm{MeV}$ and a proton chemical potential of $845.069\,\mathrm{MeV}$, at a baryon number density of $0.013 \,\mathrm{fm}^{-3}$ and a proton fraction of $0.13$. 
Heavy neutron-rich nuclei such as the hypothesized double-magic nucleus $\iso{358}{Sn}$ appear in the primordial distribution and contribute to the observed abundances after fission. 
We discuss astrophysical scenarios for the realization of this universal primordial distribution for heavy element nucleosynthesis, including supernova explosions, neutron star mergers and the inhomogeneous Big Bang. The latter scenario may be of interest in the light of early massive objects observed with the James Webb Space Telescope and opens new perspectives to explain universality of the observed $r$-process patterns and the lack of observations of population III stars.}


   \maketitle
%

 \section{Introduction}

Explaining 
the cosmic inventory of nuclear species is one of the
fundamental challenges for astrophysics and cosmology.  
Our knowledge of the isotopic abundances at various astrophysical sites has improved rapidly in recent decades thanks to a combination of advanced observational  techniques  and theoretical modeling. 
For the solar system, the isotopic abundances are well studied
\citep[e.g.][]{Asplund09,2021SSRv..217...44L,Lodders03,2009LanB...4B..712L,magg2022a}.
For other astrophysical sites, the measurements are usually restricted to elemental abundances. Particular efforts have been made to determine these abundances in metal-poor stars \citep[e.g.][]{frebel2018a, 2022ApJS..260...27R,2015ARA&A..53..631F}.

The solar abundances of isotopes (mass number $A$, charge number $Z$) are expressed in terms of mass fractions $X_{AZ}^\odot\,{=}\,A\,n_{AZ}/n_\mathrm{B}$, 
where $n_{AZ}$ is the number density of the isotope, and $n_\mathrm{B}$ the baryon number density. The mass fractions $X_{AZ}^\odot(t)$ of isotopes in the solar system are not constant but change with time $t$ due to nuclear reactions. 
 For example, fusion processes convert H to heavier elements. In the Sun, H is burned into He, which, in the future will partially be converted into C and O. For stars significantly more massive than the Sun the production of heavier elements reaches up to the iron group. 
On Earth, we observe the radioactive decay of some long-lived isotopes such as U and Th. The time-dependent distribution of elemental and isotopic abundances must have started with some initial state.
This raises the question of the origin of the elements.

The current understanding of the origin of the isotopic distribution of the elements is the standard Big-Bang nucleosynthesis model \citep[e.g.][]{2017IJMPE..2641002C,Fields:2014uja,Arcones:2022jer,2016RvMP...88a5004C}.
The expanding Universe cools down, and at the end of the ``first three minutes'' after the Big Bang, the light elements $^2$H, $^3$He, $^4$He, and $^7$Li are synthesized.
A homogeneous Universe is assumed, in which the baryon density $n_\mathrm{B}$ is constant in space.
The primordial mass fractions at the baryonic density inferred from the cosmic microwave background are theoretically determined as \citep{2017IJMPE..2641002C}
\begin{align}
 \label{hBBN}
    X^\mathrm{HBBN}_{^4\mathrm{He}}&=&0.248,\qquad X^\mathrm{HBBN}_{^2\mathrm{H}}&=&6.49\times 10^{-5},\nonumber\\
    X^\mathrm{HBBN}_{^3\mathrm{He}}&=&4.25\times 10^{-5},\qquad X^\mathrm{HBBN}_{^7\mathrm{Li}}&=&5.2\times 10^{-9}.
\end{align}
We use the notation HBBN for the homogeneous Big-Bang nucleosynthesis considered here, as we also consider in this work the inhomogeneous Big-Bang nucleosynthesis (IBBN) as an alternative scenario.
The mass fractions $X^\mathrm{HBBN}_{AZ}$ for heavier isotopes are very small, on the order of $10^{-20}$, and can be neglected.
The mass fractions $X^\mathrm{HBBN}_{^4\mathrm{He}},X^\mathrm{HBBN}_{^3\mathrm{He}},X^\mathrm{HBBN}_{^2\mathrm{H}}$ agree well with observational data  \citep{2017IJMPE..2641002C}.
One exception is $X^\mathrm{HBBN}_{^7\mathrm{Li}}$, which
is significantly higher (by a factor of ${\approx} 3.5$) than the primordial abundance derived from observations. This discrepancy is known to as the ``cosmological lithium problem''. 

The HBBN scenario has important implications. One of them is that there should be a first generation of stars -- usually referred to as population III -- that have formed from pristine HBBN material and contain virtually no metals, i.e.\ elements heavier than He. This lack of metals has implications for the formation, the evolution and, ultimately, the explosions of population III stars \citep[e.g.][]{heger2002a}.  So far, no population III stars have been found observationally. All metal-poor stars observed contain some amount of heavier elements \citep{frebel2018a,Roederer:2023spd}. Moreover, neither the pair-instability supernovae, which are supposed to result from population III stars, nor the nucleosynthetic imprints predicted for them have yet been detected.

In their seminal papers,
\citet{Burbidge:1957vc}  and \citet{Cameron_1957} proposed that all elements heavier than those produced in HBBN are formed by nuclear reactions under favorable conditions in specific
astrophysical sites. 
Over the past decades, substantial knowledge has been accumulated about the structure of nuclei and their reaction processes.
The nuclear reactions responsible for the temporal evolution of $X_{AZ}(t)$ are incorporated in nuclear reaction network (NRN) codes such as SkyNet \citep{Lippuner:2017tyn} and WinNet \citep{reichert2023a}.
Starting out from the primordial HBBN distribution $X^\mathrm{HBBN}_{AZ}$, the formation of heavier elements in suitable
astrophysical sites is modeled with these nuclear reaction networks \citep{Burbidge:1957vc,Arcones:2022jer}.
The synthesis of the nuclear species in stars up to the iron--nickel
region is well established \citep[e.g.][]{iliadis2015a}.
Our work focuses on the origin of heavy elements ($A\,{>}\,A_\mathrm{heavy}\,{\approx}\, 80$) which remained
elusive for a long time \citep{NAP10079}.

\citet{Burbidge:1957vc} invoked nucleosynthesis processes
(now called the $p$-, $s$- and $r$-processes, and some variants thereof) for their production. 
The nucleosynthesis path in the nuclear chart is assumed to run close to the neutron-drip line.
The astrophysical objects and events that provide the required conditions for the rapid neutron-capture process 
($r$-process) to operate, i.e.\ high neutron fluxes, 
turned out to be difficult to identify 
\citep{Fischer:2023ebq,Cowan:2019pkx,Wanajo:2018xex,Chen:2024gwj,Siegel:2022upa,Chen:2024acv,Arcones:2022jer}.
Most of the astrophysical sites proposed for the $r$-process involve the ejection of material from regions of high densities and contain neutron stars (NSs) or black holes (BHs). Potentially, conditions for the operation of the $r$-process are reached during core-collapse of massive stars or mergers in double-compact systems \citep{Cowan:2019pkx}.
While for many years the occurrence of the $r$-process has been associated with supernovae (SNe), where the innermost ejecta close to the central neutron star were supposed to be neutron-rich, more recent studies have cast substantial doubts on this environment for heavy-isotope production. 
Recent core-collapse supernova simulations \citep{Arcones:2012wj} indicate that  the extreme conditions necessary for the $r$-process are not reached \citep{Arcones:2012wj},
and it seems to be safe
to conclude that neutrino-driven proto-neutron star winds are excluded as
the major origin of heavy $r$-process elements \citep{Wanajo:2013hba}. 
Moreover, detailed spectroscopic observations even disfavor more extreme supernova events producing gamma-ray bursts as site for the production of $r$-process elements \citep{2024NatAs...8..774B}.

Mergers in double-compact systems containing neutron stars, black holes or white dwarfs have been investigated as alternative sites for heavy element production \citep{Chen:2024acv}. 
Spectroscopic observations of kilonova AT2017gfo associated with the neutron-star merger GW170817 confirmed the production of $r$-process elements in this event \citep{watson2019a}. 
Postprocessing the nuclear reactions in simulations of binary neutron star mergers reveals details of the production of heavy elements in such neutron-rich and dense environments \citep[e.g.][]{Fischer:2023ebq,Janka2020,Goriely2020}.
The impact of nuclear matter properties on the nucleosynthesis and the kilonova from binary neutron star merger ejecta was 
recently studied by \citet{ricigliano2024}, see also \citet{2023ApJ...951L..12J}. 
Although the production of $r$-process elements in NS mergers is confirmed by observations, it remains unclear whether they merely contribute to the enrichment of the Universe with heavy elements or whether they can account for the total cosmic abundances of these heavy nuclei. 
In particular, $r$-process elements detected in metal-poor stars require a source in the very early Universe and the time scales of enrichment with $r$-process elements due to NS mergers may be too long \citep[e.g.][]{2020ApJ...900..179K}.

An interesting finding from the observation of metal-poor stars is the similarity of the abundance pattern of heavy elements with mass numbers $A\,{\ge}\, A_\mathrm{heavy} \,{\approx}\, 80$ to that observed in the Sun \citep[e.g.][]{cowan1995a, sneden1996a, hill2002a, roederer2009a, ernandez2023a}.
This similarity initiated the notion of the ``universality of the $r$-process'' -- at least for the heaviest stable and observable elements. Such a universality seems difficult to reconcile with a production of $r$-process elements in different astrophysical sites with accidental variations in the external conditions and suggests that the observed abundance pattern reflects properties of nuclear
matter rather than conditions in the astrophysical production sites.

Together with the non-detection of population III stars, this raises the question of whether there are alternatives to the \citet{Burbidge:1957vc} scenario, which assumes the formation of the heavy elements as a product of a build-up process from light nuclei in specific astrophysical sites.
Such an alternative scenario is that of inhomogeneous Big-Bang Nucleosynthesis (IBBN), which posits that large dense objects form in the early Universe before nucleosynthesis takes place \citep{1990ApJ...362L..47T,2017IJMPE..2641003N,1987PhLB..185..281R}.
The baryonic matter expands in the space between these massive objects and forms nuclei when the density drops below the  baryon saturation density\footnote{In nuclear physics, the baryon number density $n_\mathrm{B}$ is given usually in fm$^{-3}$. 
For conversion to the mass density use 1 fm$^{-3}$ corresponds to $1.673  \,{\times}\, 10^{15}\, \mathrm{g} \, \mathrm{cm}^{-3}$. 
We consider also $T=k_\mathrm{B}T_\mathrm{therm}$ as energy which is usually measured in MeV, 1 MeV corresponds to the thermodynamic temperature $T_\mathrm{therm} = 1.1604 \times 10^{10}$ K. 
With $T_\mathrm{MeV}=T/\mathrm{MeV}$ we have $T_\mathrm{MeV}=T_9/11.604 $.}$n_\mathrm{sat} \,{=}\,0.15 \, \mathrm{fm}^{-3}$. 
Compared to HBBN, baryonic matter in this case takes a different evolutionary path in the phase diagram, and  heavy nuclei are created very early during the IBBN.
In this scenario, the homogeneous low-density state with the primordial distribution $X^\mathrm{HBBN}_{AZ}$, Eq. (\ref{hBBN}), is obsolete and is not the general initial state for the formation of elements. 
While the average baryon number density $\bar{ n}_{\mathrm B}$ in HBBN is low (of the order of $10^{-19}\,{\rm fm}^{-3}$), density fluctuations can occur, forming compact, dense objects. A reaction pathway near the neutron drip-line to form the heavy elements is not absolutely necessary, and later events such as supernova explosions
and neutron star mergers are not the only prerequisite for the formation of the heavy elements although such events can still contribute and modify the isotopic distributions.
The IBBN scenario opens new perspectives to explain universality of the observed $r$-process patterns and the lack of observations of population III stars.
Further discussion of the different scenarios for this process will be given in Sect.~\ref{sec:discussion}.

As an alternative to the nuclear reaction network calculations starting from the primordial distribution $X^\mathrm{HBBN}_{AZ}$, as given in Eq.~(\ref{hBBN}), we explore another potential pathway to the formation of heavy elements: As a starting point, we aim to find a \emph{universal
primordial state of baryonic matter in quasi-equilibrium}, 
specified by the temperature $T$, the baryon number density $n_B$, and the proton fraction $Y_p$, with a distribution of mass fractions $X_{A}^\mathrm{prim}$ of nuclei
extending out to very large values of $A$.
This state is chosen without referring to a specific astrophysical site it could be realized in. 
Given this primordial quasi-equilibrium state, we determine whether
the observed solar pattern of mass fractions $X_{AZ}^\odot$  for isotopes
with mass number $A$ and charge number $Z$ can be recovered from this state in a
freeze-out scenario with a feed-down afterburner, a reaction network which describes in particular the decay of the excited, unstable nuclei. 
In particular, we focus on the universal abundance pattern of heavy nuclei with $A \ge A_\mathrm{heavy}$.\\

The primary goal of our work is to describe expanding hot and dense nuclear matter using the freeze-out concept, which defines a primordial distribution of nuclei described by few nonequilibrium parameters related to temperature and chemical potentials. We infer such a primordial distribution for the solar abundances considering the heavy element distribution and using approaches similar to those that have proved successful in describing heavy-ion collisions (HIC).
We are concerned with the formation of nuclear clusters\footnote{We use the term ``cluster'' for referring to a nuclear particle correlation in high-density matter and not in the sense of clusters of astrophysical objects, such as stars or galaxies.} (nuclei, resonances)  in high-density matter, which requires quantum statistical approaches to describe in-medium effects, in particular Pauli blocking and the Mott effect \citep{1982NuPhA.379..536R}. These effects are not included in most recent approaches to nucleosynthesis. This freeze-out concept is of interest in any approach discussed above as site of heavy element creation.

The paper is organized as follows:
We give a short review of the behavior of expanding nuclear matter in laboratory experiments in Sect.~\ref{sec:HIC}
and discuss the gross structure of the solar mass fraction distribution in Sect.~\ref{sec:Gross}.
In Sect.~\ref{sec:M1}, we present our methods --  the Zubarev method of the nonequilibrium statistical operator, that has the freeze-out concept and the nuclear reaction kinetics as special cases -- and in Sect.~\ref{MethodII} the method of Green's functions, which describes clusters in dense matter. As results, we show in Sect.~\ref{light} the role of in-medium corrections for the light elements and in Sect.~\ref{prim} we give an estimation for a primordial distribution.
Processes leading to the final distribution of heavy nuclei are considered in Sect.~\ref{final}. In Sect.~\ref{sec:discussion}, we discuss our results and draw conclusions.

 \subsection{Analogies to nuclear cluster formation in laboratory experiments}
 \label{sec:HIC}
 
 The expansion of hot and dense nuclear matter is studied in laboratory experiments. The yields of different isotopes are observed in heavy ion collisions (HIC) at various energies. Here, we give only a few examples: At the presently highest energy densities reached in laboratory experiments, the formation of clusters  up to $^4$He ($\alpha$ particle) and its antiparticle was measured at the Large Hadron Collider (LHC) at CERN \citep{Andronic:2017pug}. 
 A laboratory test of the nuclear matter equation of state has been performed at low collision energies in the range of the Fermi energy \citep{Natowitz10,Qin:2011qp}. The production of light elements has been studied in ternary (spontaneous or induced) fission, where, for example, isotopes up to $^{24}$Mg were observed, formed in the low-density neck region at the scission point of fission of $^{241}$Pu \citep{Koster:1999hpd}.

At first glance, the distribution of observed yields of nuclei formed in HIC or ternary fission shows features very similar to that obtained from the mass action law (sometimes called the Saha equation), which describes chemical equilibrium. This concept is widely used in nuclear astrophysics. A simple statistical model -- that of nuclear statistical equilibrium \citep[NSE; see, e.g.,][]{2009ADNDT..95...96S} -- can be employed to infer the temperature and the density of a suggested freeze-out state. 

For both, modeling HIC and astrophysical processes, however, a more sophisticated approach has to account for the following problems: 
\begin{enumerate}
    \item The observed yields are not identical with the primordial yields because of afterburner processes, in particular the decay of unstable states.
    \item The primordial distribution is not a mixture of noninteracting stable nuclei. 
    We have to consider excited states and continuum correlations, but also in-medium effects such as self-energy shifts and Pauli blocking.
\end{enumerate}
While the first problem is usually taken into account in astrophysical models, the second problem is usually ignored here.

A self-consistent determination of the primordial distribution of nuclear species from observed yields is a complex problem and the use of a simple NSE model to fit the thermodynamic parameters at freeze-out is insufficient. 
Since chemical evolution is a nonequilibrium process, a quantum statistical approach is required that can describe cluster formation and the influence of the surrounding medium when expanding dense matter is considered. 
The method of the nonequilibrium statistical operator (NSO) as discussed in detail by \citet{Zubarev} represents a general approach to modeling such nonequilibrium processes, see Sect.~\ref{sec:M1}. It combines the freeze-out scenario with nuclear reaction kinetics as described by NRN codes.

For instance, it has been found \citep{Kopatch:2002bd} that in ternary spontaneous fission of $^{252}$Cf 17 \% of the observed $\alpha$ particles are produced primordially as $^5$He, which is surprising given the unstable nature of this isotope. 
In-medium effects have clearly been identified in HIC at low temperatures \citep{Qin:2011qp}.
With the approach discussed above, it was possible to determine the parameters for the primordial distribution of nuclei at freeze-out.
For HIC at ultrarelativistic energies \citep{Andronic:2017pug}, the inferred parameters at freeze-out are $T\,{=}\,157$ MeV and $\mu_B\,{=}\,0$; for HIC at energies in the Fermi energy range \citep{Qin:2011qp} temperatures of $5 - 10$ MeV and baryon number densities of $n_\mathrm{B}\,{=}\,0.02-0.04$ fm$^{-3}$ were reported, and for ternary fission of $^{242}$Pu the freeze-out values $T\,{=}\,1.29$ MeV, $n_\mathrm{B}\,{=}\,6.7 \times 10^{-5}$ fm$^{-3}$ and proton fraction $Y_p\,{=}\,0.035$ have been found \citep{Natowitz:2022npi}.
Together with an afterburner process,  that leads to the feed-down of excited nuclei and unstable states, the improved freeze-out approach proved to be very successful in explaining the yields of isotopes from expanding hot and dense nuclear matter (i.e., a fireball). With this approach, it is possible to infer quasi-equilibrium parameters for a primordial distribution in a consistent way.

In this work, we discuss the solar abundances of isotopes, in particular the heavy nuclei with mass numbers $A \ge 76$. We infer parameter values for the primordial distribution
taking into account in-medium effects, in particular self-energy effects, Pauli blocking, screening, and continuum correlations. 
 We discuss the conditions under which such nuclei are formed. 
 \citet{1987PhLB..185..281R} found parameter values of $5\,\mathrm{MeV}$ for the temperature and $0.016\,\mathrm{fm}^{-3}$ for the baryon number density. Regardless of the particular scenarios discussed above (supernova explosions, double-compact object mergers, IBBN), such conditions are of general relevance for the creation of heavy elements in the Universe.

\subsection{Distribution of the solar accumulated mass fractions}
\label{sec:Gross}

\begin{figure}[t]
  \centering
  \includegraphics[width=\linewidth]{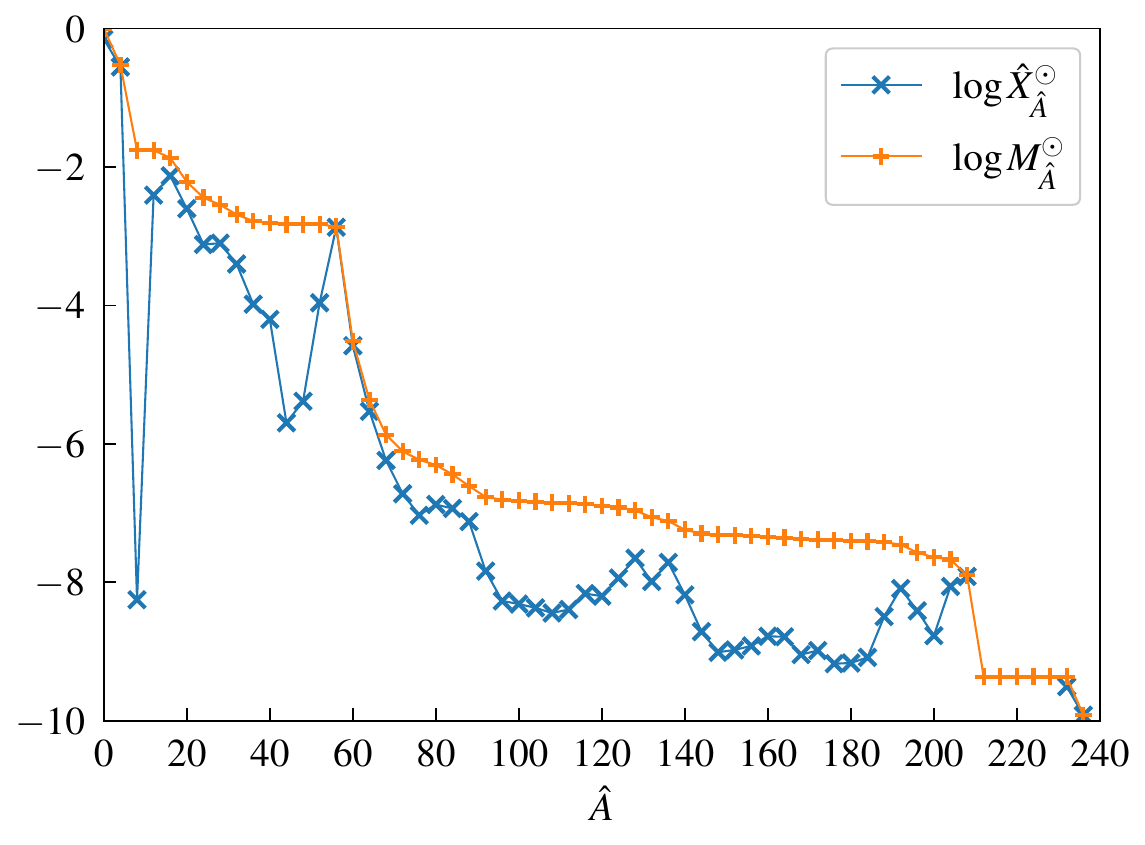}
   \caption{Accumulated mass fraction $\hat X^\odot_{\hat A}
   $, Eq.~(\ref{XA}) (blue ``$\times$'' symbols) and the $\hat A$-metallicity $M^\odot_{\hat A}
   $, Eq.~(\ref{MA}) (orange ``$+$'' symbols). Lines to guide the eyes.
   Data from \citet{Lodders03}, see Tab.~\ref{tab.1}.  \label{fig:1}
   }
\end{figure}

For our study, we take as a reference the solar abundances reported by \citet{Lodders03}. Instead of the usual detailed dependence of the abundances of isotopes on their mass number $A$ and charge number $Z$, we use the average dependence on $A$ as shown in Fig.~\ref{fig:1} introducing the accumulated mass fraction 
\begin{equation}\label{XA}
\hat X^\odot_{\hat A}=\frac{1}{n_\mathrm{B}}\sum_{A'=\hat A}^{\hat A+3} A' \sum_{Z,\nu}n_{A',Z,\nu}
\end{equation}
 and the $\hat A$-metallicity\footnote{Usually, metals ($Z$) is denoted as the set of all elements beyond He. With the $A$-metallicity we consider the fraction of material found in nuclei with mass numbers $A' \ge A$.}
 \begin{equation}\label{MA}
M^\odot_{\hat A}=\sum _{\hat A' \ge \hat A} \hat X^\odot_{\hat A'}.
\end{equation}
Here, $n_\mathrm{B}$ denotes the baryon number density, and $n_{A',Z,\nu}$ the density of clusters with mass number $A$ and charge number $Z$. The intrinsic quantum number $\nu$ denotes the excitation state of the nucleus $\{A,Z\}$.
$\hat A$ characterises the group of clusters, it can take values in $[0, 4, 8, 12, \ldots)$.
This representation of the abundance distribution of nuclear species neither shows the odd-even staggering nor the high abundances of light $n \alpha$ nuclei, which are averaged over.
The only significant deviations from the general monotonic decrease of $\hat X^\odot_{\hat A}$ with increasing $A$ as shown as red symbols in Fig.~\ref{fig:1} are the low abundance $\hat X^\odot_{8}$ related to the weakly bound  elements Li, Be, B, the large values in the maximally bound Fe--Ni region $\hat X^\odot_{56}$ (iron peak), and enhanced abundances in the regions $A\,{\approx}\, 80, 130, 195$ referred to as the first, second, and third $r$-process peak. 
These $r$-process peaks are discussed in reference to the nuclear shell model, which implies larger binding energies near the magic numbers of $Z$ and $N\,{=}\,A-Z$.

The $\hat A$-metallicity $M^\odot_{\hat A}$ shown 
in Fig.~\ref{fig:1} describes the  sum of mass fractions of isotopes with mass number $ A' \,{\ge}\, \hat A$. It is monotonically decreasing with $A$. Due to normalization, $M^\odot_{0}\,{=}\,1$. The $\hat A$-metallicity  will be of importance for our considerations, in particular to determine the freeze-out conditions. 

Using the accumulated mass fractions $\hat X_{\hat A}$ (as shown in Fig.~\ref{fig:1}) in the discussion of the time evolution of the distribution instead of the non-averaged mass fractions $X_{AZ}$ has the advantage that they are invariant with respect to special processes such as $\beta$, $\beta^+$ and $\gamma$ reactions.
Emission and absorption of neutrons or protons change the distribution $\hat X_{\hat A}(t)$ only locally ($A\,{\to}\, A\pm1$), but leave the general structure nearly unaffected, which, however, is significantly modified by $\alpha$-decay, fission and fusion processes.

Likewise, the $\hat A$-metallicity $M_{\hat A}(t)$ is invariant with respect to $\beta$, $\beta^+$, and $\gamma$ reactions. Whereas the emission or absorption of single nucleons give rise to a diffusion process (with respect to the variable $A$) for the accumulated mass fractions, the metallicity provides an integral distribution, and as such it better reflects the global changes of the mass number distribution of clusters.

In this work, we focus on the large clusters $A \ge A_\mathrm{heavy}=76$ and propose a primordial distribution different from HBBN, Eq. (\ref{hBBN}), which can explain the gross structure of the distribution of heavy element abundances. 
The formation of heavy elements is connected with a hot, neutron-rich environment, and the relaxation 
to thermodynamic equilibrium freezes out if this environment disappears. 
either the light element distribution nor the detailed isotopic distribution of the heavy elements will be considered here. 
Even-odd staggering and similar individual deviations from the gross structure can be obtained via a nuclear reaction network treatment from the primordial distribution for the expanding and cooling matter after freeze-out. The primordial distribution may be considered as the initial distribution for a NRN calculation, which describes the evolution of the distribution of nuclei after freeze-out and allows to derive the details of the final distribution.

\section{Methods}

Common approaches to describe the temporal evolution of astrophysical systems involving nuclear reactions are based on a combination of the equations of fluid dynamics coupled to a nuclear reaction network. This method is well established at low densities, where the components of the system can be considered as particles that move almost freely and occasionally have reactive collisions.
At high densities, this approach becomes invalid as in-medium effects change the properties of the components and of the reaction rates. Bound states are dissolved (Mott effect) and correlations in the continuum become important. Such effects break the assumption of an ideal fluid.
We show that the freeze-out concept, which is able to incorporate the in-medium effects, can be embedded in a general approach to non-equilibrium, and we demonstrate this general approach to in-medium corrections using a thermodynamic Green's function approach.
This allows to consistently treat matter at high densities, where the usual reaction-kinetic networks are no longer valid.

\subsection{The method of the nonequilibrium statistical operator}
\label{sec:M1}

The nonequilibrium evolution of a system is described by the statistical operator $\rho(t)$, which is the solution of the Liouville-von Neumann equation,
\begin{equation}
   \frac{\partial}{\partial t}\,\rho(t)=\frac{i}{\hbar}[H,\rho(t)],
\end{equation}
with given initial conditions.
In the Zubarev method of the nonequilibrium statistical operator (NSO) \citep{Zubarev},
the initial conditions are represented by the average values $\langle B_i \rangle^{t'}$ of a set of relevant observables $\{B_i\}$ in the past $t' \,{\le}\, t$ that characterize the state of the system. 
This information is used to construct the relevant statistical operator $\rho_\mathrm{rel}( t')$ as the maximum of the information entropy $S_\mathrm{inf}(t')\,{=}\,-\langle \rho_\mathrm{rel}(t')\rangle_\mathrm{rel}$ under given boundary conditions, i.e.\ the self-consistency conditions
\begin{equation}
  \label{selfc}
   \langle B_i \rangle^{t'} = \mathrm{Tr}\left\{\rho_\mathrm{rel}( t') B_i\right\}.
\end{equation}
As known from equilibrium statistics, these self-consistency conditions are taken into account in the variational problem via Lagrange parameters $\lambda _i(t')$, and we obtain the generalized Gibbs distribution
\begin{equation}
\label{Gibbs}
\rho_\mathrm{rel}(t')=\frac{\exp\left[-\sum_i \lambda_i(t') B_i\right]}{\mathrm{Tr}\exp\left[-\sum_i \lambda_i(t') B_i\right]}.    
\end{equation}
The Lagrange parameters $\lambda _i(t')$ must be eliminated using the  self-consistency conditions (\ref{selfc}), which represent the nonequilibrium generalizations of the equations of state.

The solution of the Liouville-von Neumann equation at given boundary conditions (\ref{selfc}) is
\begin{equation}
  \label{rhoZ}
  \rho(t)=\lim_{\epsilon \to 0} \epsilon \int_{-\infty}^{ t} \dd t' \, \mathrm{e}^{-\epsilon ( t- t')} \,
  \mathrm{e}^{-\frac{\i}{\hbar} H ( t- t')} \,\rho_\mathrm{rel}( t') \, \mathrm{e}^{\frac{\i}{\hbar} H ( t- t')}
\end{equation}
in the limit $\epsilon\,{\to}\, 0$. 

A special feature of the method of the NSO is the selection of the set of relevant observables $\{B_i\}$.
A minimum set of relevant observables are the conserved  quantities energy $H$ and the particle numbers $N_\tau$, $\tau = n, p$,
of neutrons and protons, respectively, in a volume $V$.
The solution is the generalized Gibbs distribution 
\begin{equation}
\label{Gibbseq}
     \rho_\mathrm{rel}(t') =\frac{\exp[-(H-\lambda_n(t') N_n-\lambda_p(t') N_p)/\lambda_T(t')]}{\mathrm{Tr}\exp[-(H-\lambda_n(t') N_n-\lambda_p(t') N_p)/\lambda_T(t')]} .
\end{equation}
Note that the Lagrange multipliers $\lambda_i(t')$, which generally depend on time, 
are not identical to the equilibrium parameters $T$ and  $\mu_\tau$, but can be regarded as nonequilibrium generalizations of the temperature and the chemical potentials. 
The information entropy can be unambiguously identified with the thermodynamic entropy only if the system is in thermodynamic equilibrium , and we define the quantities $T$ and $\mu_\tau$ with the known properties. 

Further correlations in the nonequilibrium state $\rho(t)$ that are not included in the set of relevant observables $\{B_i\}$ are generated dynamically and appear in $\rho(t)$ by performing the limit $\epsilon \,{\to}\, 0$. 
If each fluctuation within the relevant distribution relaxes to zero within the time $\tau_\mathrm{relax}$, the limit $\epsilon\, {\to} \,0$ in Eq. (\ref{rhoZ}) can be replaced by $\epsilon \,{\le}\, \tau_\mathrm{relax}^{-1}$. 
For $\tau_\mathrm{relax}^{-1}\,{\ge}\, \lambda_i^{-1}(t)\,\mathrm{d}\lambda_i(t)/\mathrm{d}t$, the relevant statistical operator remains almost constant during the relaxation time, so that according to Eq.~(\ref{rhoZ}) $\rho(t)$ can be approximated by $\rho_\mathrm{rel}(t)$. The memory of the system is short in this so-called Markov limit. 
All other observables are assumed to relax fast to this relevant state. Then, for instance, a Markov approximation to describe the nonequilibrium evolution is possible. 

If special fluctuations do not fulfill this requirement, they should be included in the set of relevant observables. 
This new set $\{ \tilde B_i \}$ and the corresponding relevant statistical operator $\tilde \rho_\mathrm{rel}(t)$ includes the former set of relevant observables but adds new degrees of freedom, and the temporal evolution is now described by an extended set of averages $\langle \tilde B_i \rangle^t$, for which so-called kinetic equations must be solved. 
The instant of time where the relaxation to the relevant state is no longer realized and new degrees of freedom appear marks the onset of freeze-out.
It depends on the process which is considered, and which degrees of freedom become new relevant observables, because they describe long-lived fluctuations.

If, for example, the densities of the conserved quantities are considered to be relevant observables, a hydrodynamical description with Lagrange parameters as a function of time and position is obtained \citep{Zubarev}. 
The relevant distribution $\rho_\mathrm{rel}(t)$ freezes out if the relaxation of certain fluctuations becomes too slow, so that a temporal change in the conditions (\ref{selfc}) can no longer relax to $\rho_\mathrm{rel}(t)$. 
Then these particular fluctuations should be added to the set of relevant observables $\{B_i\}$ so that a Markov approach remains possible.

To give an example from HIC, the state $\rho_\mathrm{rel}(t)$ of nuclear matter is described by the density of energy and particle density of the baryons if it is sufficiently hot and dense.
Other degrees of freedom such as composition are governed by the corresponding Lagrange parameters $T$ anf $\mu_\tau$. 
In a simple approximation, we obtain the Saha equation or the NSE. 
Better approximations lead to virial expansions and the generalized Beth-Uhlenbeck equation \citep{1982NuPhA.379..536R,Schmidt:1990oyr}.
When matter expands and cools down, inelastic, reactive collisions become rare, and the composition freezes out, also denoted as ``chemical freeze-out''.
New degrees of freedom are the partial densities of the various components, i.e.\ the nuclei. 
A NRN code describes the subsequent evolution.
Nevertheless, elastic collisions can be active for longer, so that each component approaches the equilibrium (Boltzmann) distribution as function of the center-of-mass momentum. 
If the relaxation owing to elastic collisions also becomes too slow, the occupation number in momentum space become the new degrees of freedom.
The corresponding kinetic freeze-out provides the spectrum for the various components in momentum space, and the subsequent evolution is described by the time evolution of the distribution function in momentum space.

We are interested in the reactions between the components of the expanding hot and dense nuclear matter.
When the reaction rates become too slow, the composition of the system freezes out. The composition remains at the values of the relevant state described by the parameters $\lambda_i(t)$ at freeze-out if no further reactions are considered. For the sake of simplicity, we also use these terms $T$,  $\mu_\tau$ for the nonequilibrium Lagrange parameters instead of $\lambda_T(t), \lambda_\tau(t)$ at the freeze-out time.

But even after freezing out, further reactions take place that change the abundances of the various components.
This stage of the nonequilibrium evolution due to nuclear reactions is often referred to as the afterburner processes.
The method of NSO offers the possibility of extending the set of relevant observables $\{B_i\}$. 
To derive kinetic equations, the occupation numbers of the quasiparticle states 
must be included in the set of relevant observables in order to achieve rapid convergence when calculating reaction rates \cite{Zubarev}. 
Further Lagrange parameters then appear in the generalized Gibbs distribution, such as parameters related to the single-particle distribution function. 
The Boltzmann equations result when only the occupation numbers of the components in momentum space  are considered as relevant observables. The equilibrium solution is the ideal quantum gas, without correlations and in violation of energy conservation.
To improve this, we need to add the densities of the conserved quantities as relevant observables, as known from the hydrodynamical description. 
The NSO method includes the hydrodynamic freeze-out approach and the reaction-kinetic approach using NRN codes as special cases. 

The freeze-out approach is connected with the corresponding process. 
We consider expanding hot and dense matter, where the rapid neutron exchange leads to a quasi-equilibrium, which determines also the distribution of the heavy nuclei.
If with decreasing density the absorption of neutrons by nuclei becomes slow, the primordial distribution $X_{A,Z}^\mathrm{prim}$ of heavy nuclei freezes out.
After this heavy element freeze-out (HEFO),  fission reactions between the light components can remain frequent so that the freeze-out of the composition of the light elements happens at a later time. This subsequent evolution of $X_{A,Z}^\mathrm{prim}$ after HEFO can be approximated by NRN calculations neglecting higher-order correlations, see Sect.~\ref{final}.

The aim of the present work is to infer the values of the freeze-out parameters $T, \mu_\tau$ for the generation of the solar abundances of elements,  i.e.\ the primordial distribution. 
This primordial distribution can serve as an initial condition for NRN calculations describing the evolution of the expanding hot and dense matter after freeze-out up to the observed composition. 
Before switching to the NRN approximation, however, the NSO method allows to improve the reaction-kinetic approach by taking into account in-medium effects in a systematic quantum statistical approach to obtain a consistent description of expanding hot and dense nuclear matter, which also remains valid at high densities.
In particular, the treatment of light clusters in hot and dense nuclear matter near the Mott point is a difficult task \citep{1982NuPhA.379..536R,2008PhRvC..77e5804S,2023EPJA...59..292D}. The contribution of light clusters to the equation of state, in particular the symmetry energy, is often neglected or treated in simple approximations such as the NSE \citep{1984A&A...133..175H} or the excluded volume approach \citep{Hempel:2009mc} which are not valid near the saturation density.
The need for a consistent description of nuclear matter including correlations and bound state formation in the subsaturation density range was discussed recently in context with proto-neutron stars, see, e.g., \citet{Gulminelli:2015csa,Pais:2015xoa,Pais:2019shp,Furusawa:2017auz,Furusawa:2022ktu,DinhThi:2023ioy,2023EPJA...59..292D} and references given therein.
One advantage of the NSO approach for non-equilibrium processes is that it enables the application of many-body theory, since the relevant statistical operator $\rho_\mathrm{rel}(t)$ (\ref{Gibbs}) has the form of a Gibbs ensemble.

\subsection{Green's function method}
\label{MethodII}

A main challenge is the elimination of the Lagrange parameters $T$ and $\mu_\tau$ according to Eqs.~(\ref{selfc}, \ref{Gibbseq}), i.e.\ the nonequilibrium forms of equations of state. 
Because $\rho_\mathrm{rel}(t)$ as given in Eq.~(\ref{Gibbseq}) has an exponential form, this many-body problem can be solved using the method of thermodynamic Green's functions \citep{1971qtmp.book.....F}. 
For hot and dense nuclear matter, the cluster decomposition of the single-nucleon spectral function \citep{1982NuPhA.379..536R,Ropke:1983lbc} leads to the equation of state for the total density of nucleons $n^\mathrm{total}_\tau(T,\mu_{\tau'})$ as the sum of the partial densities $n_{AZ\nu}(T,\mu_n,\mu_p)$ of the nucleus with mass number $A$, charge $Z$ and intrinsic quantum number $\nu$ (e.g.\ spin)
\begin{eqnarray}
\label{partialdens}
&&n^\mathrm{total}_n(T,\mu_n,\mu_p)=\sum_{AZ\nu} (A-Z)\,n_{AZ\nu}(T,\mu_n,\mu_p),\nonumber \\
&&n^\mathrm{total}_p(T,\mu_n,\mu_p)=\sum_{AZ\nu} Z n_{AZ\nu}(T,\mu_n,\mu_p),
\end{eqnarray}
with 
\begin{multline}
\label{nAZnu}
n_{AZ\nu}(T,\mu_n,\mu_p)=\\
\frac{1}{(2 \pi)^3}\int\limits_0^\infty \frac{\dd^3p}{\exp\left\{\dfrac{E_{AZ\nu}(p)-(A-Z)\, \mu_n-Z \mu_p}{T}\right\}-(-1)^A}.
\end{multline}
The quasiparticle energy $E_{AZ\nu}(p)$ of the nucleus depends on the center-of-mass momentum $p$.
For $A \,{\ge}\, 2$, it is the solution of the in-medium Schr{\"o}dinger equation
\begin{align}
    \label{SGl}
0=&\\
&\left[E^\mathrm{SE}(1)+\dots+E^\mathrm{SE}(A)-E_{AZ\nu}(p)\right]\psi_{AZ\nu}(p)\,(1\dots A)\nonumber \\&
+\left[1-f(1)-f(2)\right]\sum_{1'2'} V(12,1'2')\,\psi_{AZ\nu}(p)\,(1'2'\dots A)\nonumber\\&
+\mathrm{perm.},
\end{align}
which contains the self-energy (SE) term and the Pauli blocking factors $[1\,{-}\,f(1)\,{-}\,f(2)]$. 
The occupation number $f(1)$ of the single-particle state, $|1\rangle \,{=}\,|\mathbf{p}_1,\sigma_1,\tau_1 \rangle$ in momentum-spin-isospin representation,
can be approximated by the Fermi function for the uncorrelated medium.

Various approximations are known for the self-energy shift. 
We use the DD2-RMF  model \citep{Typel:2009sy} fitted to empirical data, 
\begin{align}
E^\mathrm{SE}(1)=&\sqrt{\left[mc^2+s(T,n_\mathrm{B},Y_p)\right]^2+(\hbar c p)^2}\nonumber\\
&-v(T,n_\mathrm{B},Y_p).
\end{align}
Expressions for $s(T,n_\mathrm{B},Y_p)$ and $v(T,n_\mathrm{B},Y_p)$ are given by \citet{Typel:2009sy} and 
expressions for the Pauli blocking shift are provided by \citet{2009PhRvC..79a4002R,Ropke:2011tr,Ropke:2014fia,Ropke:2020peo}.

The solution of the  in-medium Schr{\"o}dinger equation (\ref{SGl}) can have a bound part and a continuum part, which is denoted by the intrinsic quantum number $\nu$.
For $A=2$, summing over $\nu$ (including continuum correlations) leads to the Beth-Uhlenbeck formula \citep{Schmidt:1990oyr} for the second virial coefficient. 
In the non-degenerate case, we perform the integral over the center-of-mass momentum $p$ in Eq.~(\ref{nAZnu}) and obtain for the partial density $n_{AZ}(T,\mu_n,\mu_p)$ of the nucleus in the channel $\{A,Z\}$ the expression
\begin{multline}
\label{nAZ}
    n_{AZ}(T,\mu_n,\mu_p)=\\
    R_{AZ}(T,\mu_n,\mu_p)\, \left( \frac{2 \pi \hbar^2}{AmT}\right)^{-3/2}\\ \times \exp\left\{-\frac{E^0_{AZ}(T,\mu_n,\mu_p)-(A-Z)\, \mu_n-Z \mu_p}{T}\right\},
\end{multline}
where $E^0_{AZ}$ is the medium-modified ground state energy of the nucleus $\{A,Z\}$. 
The degeneracy factor $g_{AZ}$ and
the sum over all excited states, including the continuum contributions, are summarized in the prefactor  $R_{AZ}(T,\mu_n,\mu_p)$, the intrinsic partition function. For the light elements, the excited states of the nuclei and their degeneracy are known \citep{nuclei} 
so that the summation can be performed within the intrinsic partition function and the continuum contribution to the virial form  \citep{Ropke:2020hbm,Natowitz:2022npi}. For the heavier nuclei,
the summation over their excited states can be replaced by the integral over the density of states \citep{BohrM} 
\begin{multline}
R_{AZ}=g_{AZ} \\ +\frac{1}{12} \left(\frac{15 \pi^2}{A}\right)^{1/4}\int\limits_{E_\mathrm{min}}^{E_\mathrm{max}} E^{-5/4}dE \exp\left\{\frac{2\sqrt{E\, A/15}-E}{T}\right\},
\end{multline}
where we take $E_\mathrm{min}\,{=}\,25\,\mathrm{MeV}/A$ and $E_\mathrm{max} $ as the binding energy of the bound state $\{A,Z\}$. For a more detailed discussion of the intrinsic partition function see \citet{Rauscher:2003ti}.

From the Beth-Uhlenbeck results for the virial coefficients for $^2$H, $^4$H, $^5$He and $^8$Be, the approximation
\begin{multline}
R^\mathrm{virial}=\left[\exp\left(-\frac{E_{AZ}^\mathrm{thres}/\mathrm{MeV}+1.129}{0.204}\right)+1\right]^{-1} \\ 
\times \left[\exp\left(-\frac{E_{AZ}^\mathrm{thres}/\mathrm{MeV}+2.45}{T_\mathrm{MeV}}\right)+1\right]^{-1}
\end{multline}
was suggested for the light clusters by \citep{Ropke:2020hbm}, with the energy $E_{AZ}^\mathrm{thres}$ of the continuum edge of scattering states (for resonances the negative value of the resonance energy).
The Pauli blocking shifts for light elements $A \,{\le}\, 16$  were considered in \citet{Ropke:2020peo}. 
For $10 \,{\le}\, A \,{\le}\, 16$, the expression 
\begin{equation}
\Delta E_{AZ}^\mathrm{Pauli}\approx 1064\,\mathrm{e}^{-0.05103\, T_{\rm MeV}}\,(N n_n+Z n_p)\,{\rm MeV \,fm}^3 
\end{equation}
was found.
A more detailed treatment requires to take into account not only the influence of $T$ but also the dependence on the center-of-mass momentum of the nucleus. Both effects result in a decrease of the Pauli-blocking shift. 
A further effect is that the bound-state contribution to the partial densities (\ref{partialdens}) disappears if the binding energy becomes zero, but, according to the Levinson theorem, a contribution of the continuum appears, which can be interpreted as a relic of the merging bound state with the continuum \citep{Schmidt:1990oyr}.
We impose a Pauli blocking shift
\begin{equation}
\label{Paulidelta}
    \Delta E_{AZ}^\mathrm{Pauli}(T,n_\mathrm{B},Y_p)= A \delta_{AZ}^{\rm Pauli}(T,Y_p) n_\mathrm{B}
\end{equation}
for $Z\,{=}\,Y_p\, A$, which describes the effective suppression of the contribution of the nucleus $\{A,Z\}$ to the partial density.

The solution of the in-medium Schr{\"o}dinger equation (\ref{SGl}) for the ground state (and excited states) of nuclei is very complex and requires an expression for the interaction potential. Instead, the empirical data for the binding energies can be used, which are known for many isotopes (\citealp{nuclei}, 
see also \citealp{Wang:2021xhn}). Since we need values for the binding energies of isotopes far from stability, we estimate them using model calculations. 
Hartree-Fock-Bogoliubov calculations \citep{2009PhRvL.102o2503G}, microscopic mass formulas \citep{1995PhRvC..52...23D}, and calculations with the  Finite-Range Droplet Model \citep{2016ADNDT.109....1M} were proposed for this purpose. 
Since we are not dealing with the fine structure of element abundances, we use simple approaches for the general behavior of heavy nuclei.
Here, we consider the liquid droplet model (LDM) for the ground-state bound state energies $E^0_{AZ}$ of Eq.~(\ref{nAZ}),
\begin{equation}
\label{BW}
E^{0,\rm LDM}_{A,Z}=-a_\mathrm{b} A+a_\mathrm{s} A^{2/3}+a_\mathrm{c}  \frac{Z(Z-1)}{A^{1/3}}+a_\mathrm{a} \frac{(A-2Z)^2}{A}.
\end{equation}
The parameters $a_\mathrm{b}\,{=}\,15.76 \,\mathrm{MeV}$, $a_\mathrm{s}\,{=}\,17.81\,\mathrm{MeV}$, $a_\mathrm{c}\,{=}\,0.714\,\mathrm{MeV}$ and $a_\mathrm{a}\,{=}\,23.7\,\mathrm{MeV}$ are known for isolated clusters \citep{rohlf1994a}. 
Further contributions to the LDM consider pairing (not of relevance for the averages $\hat X_{\hat A}$) and shell corrections \citep{1995PhRvC..52...23D,2016ADNDT.109....1M,Royer2008a,2009EPJA...42..269D}, but see also Skyrme--Hartree-Fock--Bogoliubov (SHFB) calculations \citep{2009PhRvL.102o2503G} and machine learning approaches \citep{2024arXiv240411477M}.
We use the parametrization of the shell correction according to \citet{2009EPJA...42..269D}. The shell correction to the bound state energy is
\begin{equation}
\label{shellcorr}
    E_\mathrm{shell}(N,Z)/\mathrm{MeV}=-1.39 S_2+0.02 S_2^2+0.003S_3+0.075S_{np},
\end{equation}
where
\begin{eqnarray}
    &&S_2=\frac{n_v(D_n-n_v)}{D_n}+\frac{z_v(D_z-z_v)}{D_z}, \nonumber\\
    &&S_3=\frac{n_v(D_n-n_v)(2n_v-D_n)}{D_n}+\frac{z_v(D_z-z_v)(2z_v-D_z)}{D_z}, \nonumber\\
     &&S_{np}=\frac{n_v(D_n-n_v)}{D_n}\frac{z_v(D_z-z_v)}{D_z},
\end{eqnarray}
with $n_v (z_v)$ denoting the number of valence neutrons (protons) and $D_n (D_p)$ the degeneracy of the neutron (proton) valence shell, e.g., $D=32$ for the 50-82 shell. We use the magic numbers 2, 8, 14, 20, 28, 50, 82, 126, 184 and add the numbers 228 and 308 given by \citet{2013JPSJ...82a4201K}.

In addition to the self-energy and Pauli blocking associated with the strong interaction, another in-medium correction is the screening of the Coulomb interaction.
Instead of more sophisticated approaches for the dielectric function of a charged particle system using Green's function techniques \citep{KKER},
we use the simple Wigner-Seitz model, in which the Coulomb part of the binding energy, Eq.~(\ref{BW}), of a droplet cluster is reduced so that
\begin{equation}
\label{screen}
 a^\mathrm{matter}_\mathrm{c}=0.714 \left[1-1.5 \left(\frac{Y_p \,n_\mathrm{B}}{n^0_p}\right)^{1/3}+\,0.5 \,\frac{Y_p\, n_\mathrm{B}}{n^0_p}\right]\,\mathrm{MeV},
\end{equation}
with $n_p\,{=}\,Y_p \,n_\mathrm{B}$ as the proton density of the matter and $n^0_p$ as the proton density in the cluster.

The method of Green's functions offers a systematic approach to treat in-medium effects, in particular self-energy and Pauli blocking.
At high densities, nuclear matter cannot considered as ideal mixture of baryons and nuclei. 
A semiempirical approach to account for the effect of Pauli blocking is the concept of an excluded volume, see, e.g., \citet{Hempel:2009mc}.
This concept neither is able to reproduce details of the Pauli blocking such as momentum and temperature dependence, nor the dissolution of bound states (Mott effect).

\subsection{Freeze-out approach for laboratory experiments}
\label{Labor}
We demonstrate the freeze-out concept and the afterburner approach for the formation of the light elements H, He from expanding hot and dense nuclear matter.
To explain the evolution of the primordial distribution after freeze-out, we consider nuclear matter with only two elements, H and He, together with neutrons in the primordial distribution of isotopes. 
Can we derive the Lagrange parameters $T$ and  $\mu_\tau$ at freeze-out from observed yields of the stable isotopes $^1$H, $^2$H, $^3$He, $^4$He?
This problem was investigated for HIC \citep{Qin:2011qp,Natowitz:2010ti} and ternary fission \citep{Ropke:2020hbm,Natowitz:2022npi}. 
The primordial distribution contains not only the stable isotopes but also all unstable isotopes, including their excited states and continuum contributions. The energies of the ground state and the excited states are modified by in-medium effects such as self-energy and Pauli blocking. A simple NSE description is invalid because it neglects these effects.
For HIC at collision energies of $35\,\mathrm{MeV}$ per nucleon, \citet{Natowitz:2010ti} reported the freeze-out baryon density $n_\mathrm{B} \,{=}\, 0.005 \, \mathrm{fm}^{-3}$ for $T \,{=}\, 5 \,\mathrm{MeV}$ and a proton fraction of $Y_p=0.41$. 

In ternary fission, nuclei are formed during scission in the neck region, where the proton fraction $Y_p$ is low, so that neutron-rich isotopes appear more abundantly.
For instance, in addition to the stable nuclei $^2$H, $^3$H, $^3$He, $^4$He, the abundance of unstable nuclei such as $^5\mathrm{He}$ in the hot and dense primordial matter is calculated, which decay after freeze-out and feed the observed yields of $^4\mathrm{He}$. 
Their contribution of about 17 \% to all observed $^4\mathrm{He}$ events was determined by experiments \citep{Kopatch:2002bd}, in which the $n-\alpha$ correlations in the ternary spontaneous fission of $^{252}\mathrm{Cf}$ were measured. 
The same was observed for the unstable isotope $^7\mathrm{He}$, which feeds the final yield of $^6\mathrm{He}$. Also $^4\mathrm{H}$ is obtained in the primordial distribution, but it feeds the final yield of $^3\mathrm{H}$. 
If all decay processes are taken into account \citep[see][]{Ropke:2020hbm,Natowitz:2022npi}, the primordial distribution at freeze-out converts to the final distribution. This analysis allows to infer the Lagrange parameters for the primordial distribution (\ref{nAZ}) from the observed elemental and isotopic abundances.
Together with the observed yields of $^2$H 
and $^3$H, a freeze-out parametrisation for $^{241}$Pu of the neck region at fission was obtained with $T\,{=}\,1.29\, \mathrm{MeV}$, baryon density $n_\mathrm{B}\,{=}\,6 \times 10^{-5}$ fm$^{-3}$ and proton fraction $Y_p\,{=}\,0. 035$. 
For this purpose, a least-squares fit was performed between the observed yields and the calculated final distribution, which is determined by the Lagrange parameters $T$ and $\mu_\tau$. 

 Comparing the calculated yields of $^6\mathrm{He}$ and $^8\mathrm{He}$ with the observed yields, the suppression of weakly bound clusters due to Pauli blocking was discussed \citep{Ropke:2020hbm}. These in-medium effects are more evident in the suppression of the weakly bound isotopes $^{11}\mathrm{Li}$ and $^{19}\mathrm{C}$ \citep{Natowitz:2022npi}. 
 Medium effects are also evident in HIC \citep{Qin:2011qp}.

\section{Results}
To determine the parameter values for the  primordial distribution, we consider the heavier elements which are almost not influenced by stellar burning  cycles.
We discuss the heavy element distribution $A \,{\ge}\, 76$ of the solar abundances as shown in Fig.~\ref{fig:1}.

When hot and dense matter expands and cools down, after freeze-out the primordial distribution  of the heavy elements evolves by decay processes such as particle emission, $\alpha$ decay, and fission.  The reverse reactions are progressively suppressed  with decreasing densities. 
The distribution of the heavy-element abundances shifts towards lower $A$. 
We focus here  on the expansion of hot and dense matter, independently on the path for reaching this state. We intend to determine the Lagrange parameters $T$ and $\mu_\tau$, which describe the relevant statistical operator (\ref{Gibbseq}). These parameters may be considered as the nonequilibrium generalizations of temperature and chemical potentials of a local thermodynamic equilibrium distribution.
We demonstrate this freeze-out approach which is known from nuclear reaction in Sect.~\ref{light}, discuss the decay processes in Sect.~\ref{final} and present an estimate of the Lagrange parameters $T$, $\mu_n$ and $\mu_p$ at HEFO in Sect.~\ref{prim}. 

\subsection{Primordial distribution for the light elements H, He}
\label{light}
 As discussed in Sect.~\ref{Labor}, we follow an approach similar to that taken in interpreting ternary fission and HIC experiments. 
 To move on to astrophysical conditions, we start out with calculations for a nuclear system with $Z\,{\le}\, 2$. 
Considering a final mass fraction $X_{^4\mathrm {He}}
\,{=}\,0.245$ of $^4\mathrm{He}$ in the H--He system, which is close to the observed solar composition, we  search for possible primordial distributions characterised by $T$ and $n_\mathrm{B}$, taking the proton fraction $Y_p$ as a free parameter. 
We include all isotopes of H and He, as well as their excited states, in the primordial distribution. 
Continuum correlations and in-medium modifications are taken into account. 
We assume that
in the expansion phase after freeze-out, neutrons and all H isotopes decay to the stable nuclei $^1$H and $^2$H; they feed-down to these stable isotopes. 
Similarly, all He isotopes decay to the stable nuclei $^3$He and $^4$He, they feed-down to these stable He isotopes. The remaining neutrons feed the yield of $^1$H.

\begin{figure}[ht]
  \centering
  \includegraphics[width=\linewidth]{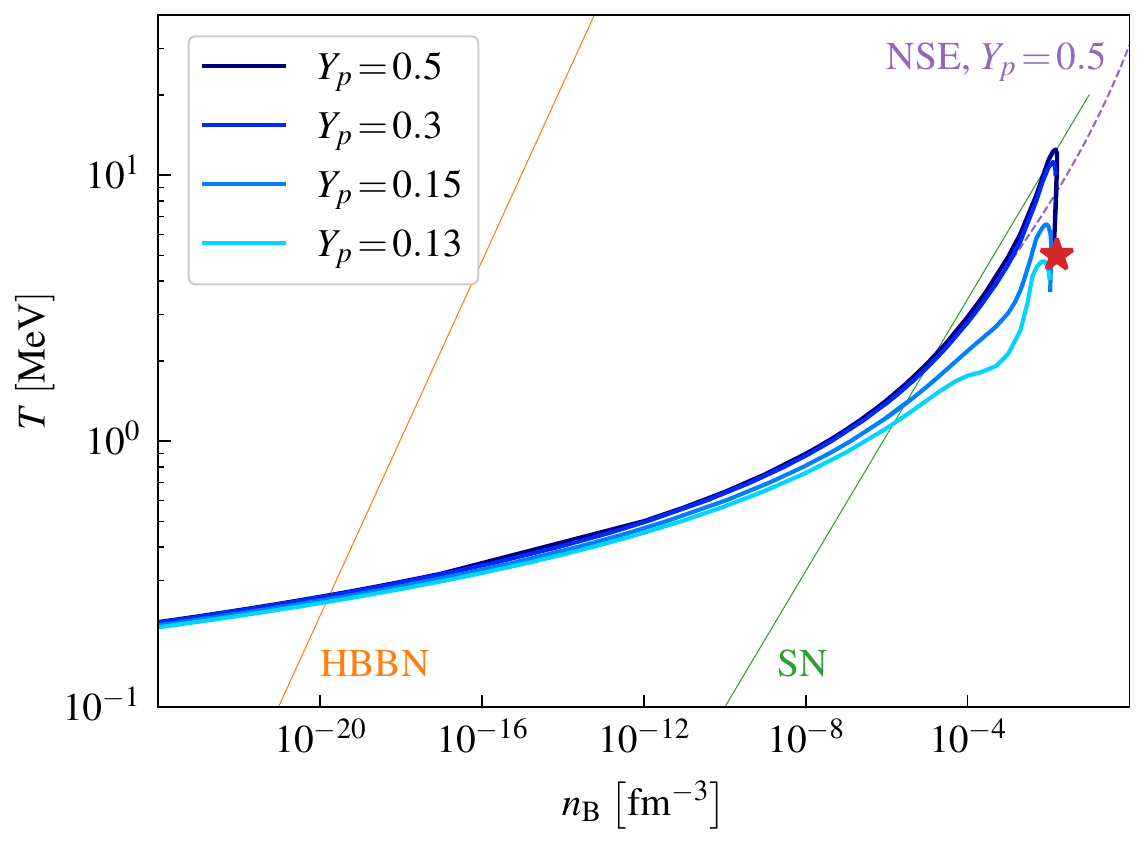}
  \includegraphics[width=\linewidth]{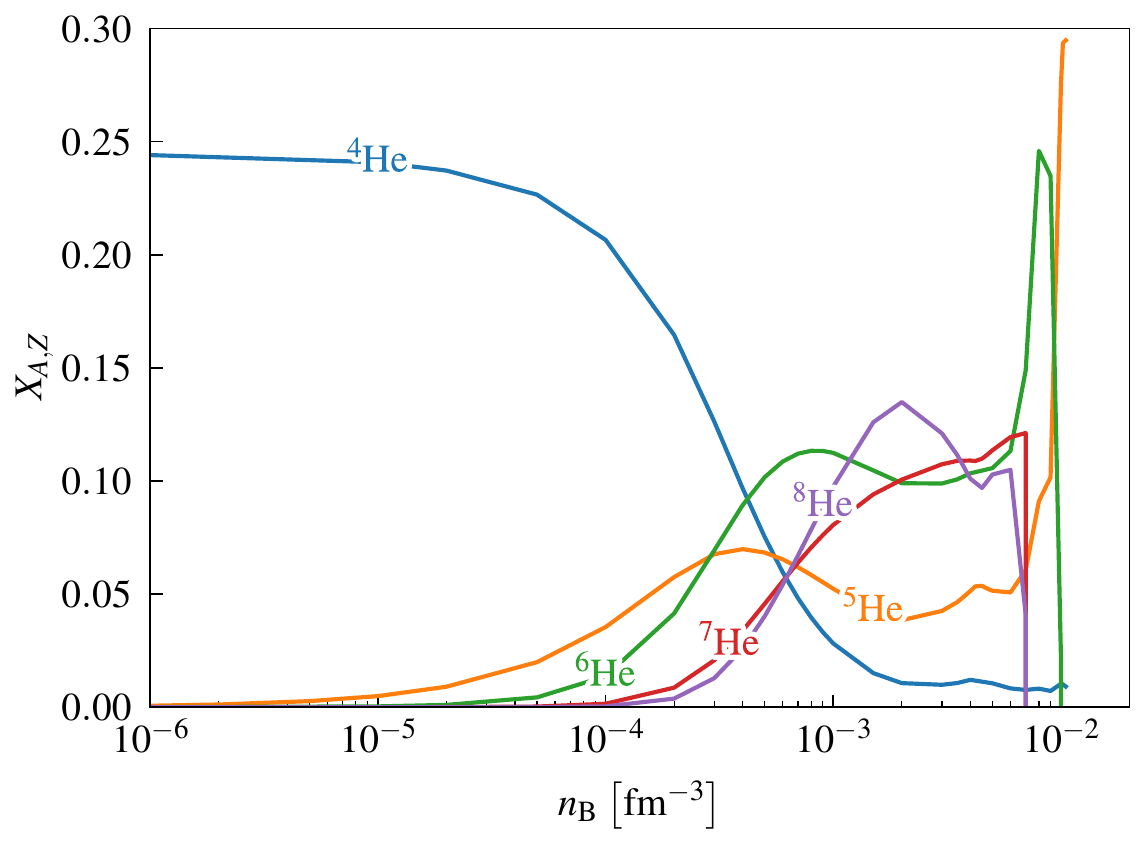}
   \caption{\label{fig:varsize} Panel ({\bf a}): Nuclear matter phase diagram. 
   For given proton fraction $Y_p = 0.5, 0.3, 0.15$, and $0.13$, the temperature is shown at which the final mass fraction of $^4$He takes the value 0.245,  as function of the baryon density $n_\mathrm{B}$ (full lines). 
   Only light elements $Z\le 2$ are considered. For comparison, the ideal gas (NSE) approximation (dashed) for $Y_p = 0.5$ is also shown.
   The red star denotes the estimated freeze-out parameter values of  \citet{1987PhLB..185..281R}. 
The dashed line indicates the pathway of the homogeneous big bang path (HBBN), whereas the dotted line denotes a typical path for supernova explosions (SN) \citep{2014EPJA...50...46F}.\\
Panel ({\bf b}): Composition at freeze-out for the final distribution with $X_{^4\mathrm{He}}=0.245$, $Y_p=0.13$. Temperatures as function of baryon density $n_\mathrm{B}$ according Panel (a).  The primary mass fractions of various He-isotopes $^4$He - $^8$He are presented. }
\end{figure}

The results of the calculations for $Y_p\,{=}\,0.5, 0.3, 0.15$ and $0.13$ are shown in Fig.~\ref{fig:varsize}. 
For these different parameter values of the proton fraction $Y_p$, the parameter values $(T, n_\mathrm{B})$ are shown for which a mass fraction $X_{^4\mathrm{He}}\,{=}\,0.245$ of $^4$He is recovered in the final distribution. The corresponding primordial distributions also contain isotopes of H and He with higher mass numbers. In particular, in the neutron-rich case $Y_p\,{=}\,0.13$, the isotopes $^5$He and $^6$He will be abundant with increasing density and even exceed the primordial yield of $^4$He. 
The mass fraction for the heavy helium isotopes $^5$He, $^6$He, etc.\ increase with higher densities. Note that the mass fraction 0.245 refers to the final $^4$He components given by $X^\mathrm{final}_\alpha\,{=}\,(4/n_\mathrm{B})\sum_{A\ge 4} n_{A,2}$. The emitted neutrons from the heavier He isotopes eventually contribute to the stable components of H. 
 
However,  above a critical density of about $n_\mathrm{B}\,{=}\,0.011 \, \mathrm{fm}^{-3}$ the helium clusters are dissolved due to Pauli blocking and the Mott effect, see Fig.~\ref{fig:varsize}.
For comparison, the corresponding curve at $Y_p\,{=}\,0.5$ for the simple NSE considering the stable isotopes of H, He is also shown in the figure. 
 The deviation at high densities is caused by the account of unstable states of helium isotopes. 
 A similar behavior is encountered for the H isotopes. With increasing density, the abundance of heavier H isotopes increases, but it goes down at the Mott density, where the bound clusters are dissolved. However, the protons ($^1$H) and the neutrons remain in the nuclear matter as quasiparticles.
Pauli blocking, which results in the lowering of the binding energies and finally gives rise to the dissolution of bound states, is a dominant effect at subsaturation densities $10^{-3}\,\mathrm{fm}^{-3}\,{\le}\, n_\mathrm{B} \,{\le}\, 0.15$ fm$^{-3}$. 
 We consider $n_\mathrm{B}\,{=}\,0.011 \, \mathrm{fm}^{-3}$ as a critical baryon density at which bound states are formed. This value depends only weakly on the proton fraction $Y_p$.
 It should be mentioned that a more detailed treatment of correlations in nuclear matter predicts a smoother dissolution of the bound states if the dependence on the center-of-mass momentum and the contribution of the continuum are taken into account rigorously.

Comparing the two cases shown in Fig.~\ref{fig:varsize} -- the quantum-statistical calculation with in-medium corrections  (solid line) and the NSE (dashed line) -- one main difference is the suppression of cluster formation at increasing density. 
All bound states are dissolved near $n_\mathrm{B}\,=\, 0.1$ fm$^{-3}$.
  The kink behavior near $n_\mathrm{B}=0.005$ fm$^{-3}$ is related to the disappearance of $^9$He and $^{10}$He; they are not shown here, since their abundances are small.

In-medium corrections, in particular Pauli blocking and the Mott effect, are important phenomena when calculating the primordial distribution for HIC or ternary fission experiments.
Since we expect the site of heavy element formation to fall in the high-density region, we must also take these effects into account. A description in terms of NSE is not valid in this regime.

\subsection{Heavy element chemical freeze-out and the final abundances of isotopes}
\label{final}

As discussed in Sec. \ref{sec:M1}, the freeze-out concept defines the primordial distribution, which is the relevant distribution at the instant of time  when the fluctuations no longer relax fast enough to restore this relevant state. 
The mass fractions $X_{A,Z}(t)$ of the various components remain in local thermodynamic equilibrium up to the heavy element freeze-out (HEFO) 
where the primordial distribution $X_{A,Z}^{\rm prim}$ is obtained.
The balance of emission and absorption of neutrons by nuclei necessary for the chemical equilibrium of heavy nuclei $A \,{\le}\, A_\mathrm{heavy}$ 
is not realised if the density of the neutron environment decreases. 
After HEFO, we treat the evolution of the distribution $X_{A,Z}(t)$ as an approximation similar to the case of HIC and ternary fission (see Sect.~\ref{Labor}) considering the decay of excited states as feed-down processes.

In this reaction-kinetic stage of the evolution, 
$\beta$-and $\gamma$ decays transform the excited nuclei to lower-energy states without changing their mass number $A$.
Only processes that change the mass number $A$ are of relevance for the evolution of $\hat X_{\hat A}(t)$. 
For the heavy nuclei, the feed-down to other mass numbers happens if nucleons are emitted  (for instance due to the evaporation of neutrons), $\alpha$ decay, and fission.

Excited neutron-rich nuclei emit neutrons. This evaporation process is well known, for instance at fission.
There, daughter nuclei are formed, which emit the average number $\bar \nu$ of neutrons with an energy distribution corresponding to a temperature of about $1\,\mathrm{MeV}$.
The average neutron multiplicity is nearly proportional to the excitation energy, $\bar \nu \,{=}\, a E_\mathrm{ex}+b$, with $a_{^{235}\mathrm{U}}\,{=}\,0.12\,\mathrm{MeV}^{-1}$, $a_{^{239}\mathrm{Pu}}\,{=}\,0.14\, \mathrm{MeV}^{-1}$, and $b \,{\approx}\, 2.5$ \citep{Fraisse:2023hfx}.
At temperature $T$, a nucleus with $A$ nucleons has an excitation energy of $E_\mathrm{ex}\,{=}\,A\, T^2/c$ with
$c\,{=}\,8 \,\mathrm{MeV}$ as obtained from the phenomenological Fermi gas approximation \citep{PhysRevC.103.025806} and an empirical value of
$c \,{\approx}\, 10\,\mathrm{MeV}$ \citep{Pochodzalla:1995xy}.
We estimate the neutron multiplicity evaporated from a nucleus $A$ at temperature $T$ as
\begin{equation}
\label{evap}
    \bar \nu(A,T)= a \frac{T^2}{c} A = d\, A\, T^2_\mathrm{MeV}
\end{equation}
with $d\,{\approx}\, 0.012$. The increase with $A$ is also shown in the evaporated neutron distribution of spontaneous fission of $^{252}$Cf \citep{PhysRevC.103.025806}.

The emission of $\bar \nu$ neutrons (\ref{evap}) from a nucleus with mass number $A$ gives a nucleus with average mass number $A'\,{=}\,\gamma A$, with $\gamma\,{=}\,1-\bar \nu/A$ (for averages, we consider $A$ as continuous variable, to be replaced by discrete values with corresponding probabilities 
 at the end). This process of evaporation transforms the primordial distribution $\hat X_{\hat A}^\mathrm{prim}$ to $\hat X_{\hat A}'$.
The mass fraction of the interval $[A,A+3]$ is reduced by $\gamma$, but also the interval is changed by the same value so that the  with the 
values $\hat X_{\hat A}^\mathrm{prim}$ at new position $\gamma \hat A$ remain, $X'(A)\,{=}\,X^{\rm prim}(A/\gamma)$ taking $A$ as a variable. 
Since $A$ is a discrete number, we have to redistribute the mass fraction over the intervals of $\hat A$ in order to obtain the accumulated mass fraction after evaporation, but the height remains almost unchanged. The metallicity is shifted and reduced, $M'(A)\,{=}\,\gamma M^{\rm prim}(A/\gamma)$ taking $A$ as variable.

The primordial distribution $\hat X_{\hat A}^\mathrm{prim}$ is characterized by the parameters $T$, $\mu_n$ and $\mu_p$, and it extends over arbitrary $A$,
including values of the order of $10^3$. 
The heaviest stable nucleus is $^{208}$Pb. Nuclei with $A \,{>}\, 208$ will decay.
They feed down lower values of $A$ in the final distribution.
The processes of main interest are $\alpha$ decay, binary fission, ternary fission and multifragmentation.

The $\alpha$ decay of the actinides and superheavy elements is well known.
Chains of $\alpha$ decay end in the lead region.
The peaks near the magic numbers (132, 208, etc) can be related to these chains of $\alpha$ decays.
Part of nuclei with $A \,{>}\, 208$ perform $\alpha$ decay after HEFO and feed down to the nuclei found in the intervals belonging to the accumulated mass numbers  $\hat A\,{=}\,204$ and $208$.
A more detailed treatment of this afterbuner process should take into account the half-lives for the $\alpha$ decay, which are below $1 \, \mathrm{s}$ for $A \,{>}\,290$. 
The isotopes $^{232}$Th, $^{238}$U and $^{244}$Pu have extremely long half-lives and are present in the solar abundances.

Fission is the dominant decay mode of nuclei in the ground state with $Z \,{\le}\, 104, N \,{\ge}\, 158$.
It becomes more important for the decay of excited nuclei.
The role of fission for nucleosynthesis became of increasing interest during the last decade 
\citep{panov2018a,Panov:2023tfn,Chen_2023,PhysRevC.103.025806,Eichler:2014kma,2008PhRvC..77c5804B,2024A&A...688A.123X}.
Since we are not concerned with the problem how fission becomes active starting from low-density matter, the primordial distribution at HEFO contains all possible nuclei and resonances so that only the decay of them is of interest. 
Although nuclear fission is of great importance for the description of the abundances of the heavy elements, only limited experimental data are available.
As described in \citep{PhysRevC.103.025806}, experimental data and the GEF (general description of fission observables) calculations tend to indicate symmetrical fission for large mass numbers $A$.
Fission fragments populate the regions near $A\,{=}\,116$ $(Z\,{=}\,50)$ and $A\,{=}\,176$ $(Z\,{=}\,70)$.
Part of mass fraction observed in that range can be attributed to a fission origin \citep{Roederer:2023spd}.

After HEFO, the evolution of the distribution $X_{A,Z}(t)$ can be approximated by NRN calculations as afterburner, based on detailed information about the isotopes, their excited states and reaction cross sections.
This is not the aim of the present work.
We are interested to infer the parameter values $T$, $\mu_n$ and $\mu_p$ for a primordial distribution.
Therefore, we consider only the gross structure of the distribution function given by the accumulated mass fractions $\hat X_{\hat A}$.
A  model for the individual isotopic abundances requires a more detailed approach such as NRN codes.
Instead, we only discuss the afterburner evolution in general terms, which connects the primordial distribution at HEFO
with the accumulated mass fractions observed from the solar abundances.

For future work, NRN calculations would be of interest to determine the final distribution $X_{A,Z}$.
It should be mentioned, however, that these calculations also contain some approximations if HEFO appears at extreme conditions.
The typical NRN approximation neglects in-medium corrections such as Pauli blocking and self-energy effects, which are included in the relevant statistical operator $\rho_\mathrm{rel}$ at freeze-out, which also describes correlations in the continuum. 
For instance, already the modeling of the afterburner processes for ternary fission \citep{Ropke:2020hbm,Natowitz:2022npi} requires information about excited states of light nuclei and branching rations for decay processes in a dense medium. These are known only in some approximation.

\subsection{Primordial distribution from the heavy elements}
\label{prim}

To construct the primordial distribution, the binding energies of nuclei are needed. 
As generally accepted, the formation of the heavy elements  requires a neutron-rich, hot environment; see also 
Fig.~\ref{fig:varsize}. Neutron-rich nuclei are formed, with binding energies not known from experiments. 

Hartree-Fock-Bogoliubov or Finite-Range Droplet Model calculations are applied to determine the binding energies of nuclei far from stability,
but one has to consider the effects of the medium which influence the properties of nuclei. 
For the light nuclei, where the binding energies are measured experimentally, the shifts of the binding energies due to self-energy and Pauli blocking are known.
For the heavy nuclei, we use the Liquid Droplet Model (LDM) (\ref{BW}). 
Its parameter values $a_{\rm b}$, $a_{\rm s}$, $a_{\rm c}$, and $a_{\rm a}$ depend on the surrounding medium. 
Within the Wigner-Seitz approximation, the modification of the Coulomb term $a_{\rm c}$ was already discussed in Sect.~\ref{MethodII}. 
Because of screening and a low proton fraction, the Coulomb term becomes very small at high densities.
We assume that the asymmetry energy $a_{\rm a}$ is not modified. 

Self-energy shifts affect bound states as well as continuum states so that it effectively modifies the chemical potentials.
We use the DD2-RMF result \citep{Typel:2009sy} given in Sect.~\ref{MethodII} for the nucleons in the nuclei, which leads to a renormalization of the chemical potentials $\mu_n$ and $\mu_p$. 
Typical values for the self-energy of nucleons at $T\,{=}\,5\, \mathrm{MeV}$, $n_\mathrm{B}\,{=}\,0.01\, \mathrm{fm}^{-3}$ and $Y_p\,{=}\,0.13$ are $\Delta E^{\rm SE}_n\,{=}\,-6.31\, \mathrm{MeV}$ and $\Delta E^{\rm SE}_p\,{=}\,-13.08\, \mathrm{MeV}$.
The Pauli blocking compensates partially the self-energy shift and leads to the dissolution of the bound states (Mott effect), as discussed in Sect.~\ref{MethodII}.
According to Eq.~(\ref{Paulidelta}), we introduce the parameter $\delta^{\rm Pauli}(T,Y_p)$ as an average value of $\delta_{AZ}^{\rm Pauli}(T,Y_p)$ and add this to the bulk parameter $a^\mathrm{matter}_{\rm b}(T,n_{\rm B},Y_p)\,{=}\,a_{\rm b}-\delta^{\rm Pauli}(T,Y_p) n_{\rm B}$. 
For $T\,{=}\,5\, \mathrm{MeV}$ and $Y_p=0.13$, we estimate $a^\mathrm{matter}_{\rm b}(T,n_{\rm B},Y_p)\,{\approx}\, 14.26\, \mathrm{MeV}$.

\begin{figure}[t]
  \centering
  \includegraphics[width=\linewidth]{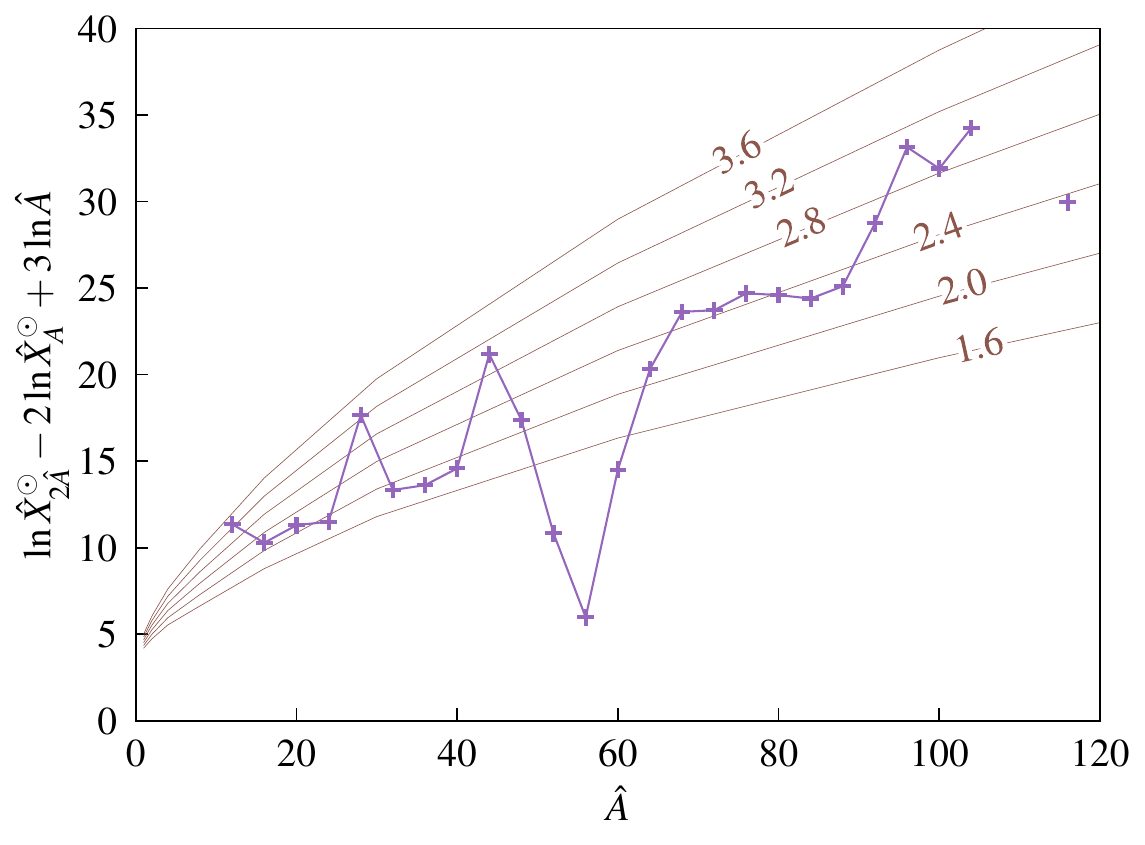}
   \caption{
   Double ratio of accumulated mass fractions. 
   Solar distribution $\ln \hat X^\odot_{2 \hat A} - 2 \ln{\hat X}^\odot_A + 3 \ln \hat A$ 
   (purple ``$+$'' symbols, lines to guide the eyes; values from Table~\ref{tab.1}) 
   compared with liquid droplet model calculations, Eq.~(\ref{doubleratios}), brown curves, for various parameter values $a_\mathrm{s}/T$ as given in the line labels. 
   Further parameter values are provided in the text.}
   \label{fig:ratios}
\end{figure}

For an estimate of the parameter value of $a^\mathrm{matter}_\mathrm{s}$
, we consider the ratio $\hat X_{2A}/\hat X^2_A$ as shown in Fig.~\ref{fig:ratios}. Using the solar data (see Tab.~\ref{tab.1}) we expect a smooth average dependence on $A$. However, there are strong deviations related to the peaks in $\hat X^\odot_{\hat A}$ owing to the magic numbers and the iron peak (see Fig.~\ref{fig:1}), which are not relevant for our present discussion of the LDM.

We employ the LDM (\ref{BW}) to calculate the ratio $\hat X_{2A}/\hat X^2_A$. 
For a nucleus with mass number $A$ we assume an average charge number $\bar Z(A)\,{=}\,Y_p A$.
From Eq.~(\ref{nAZ}) we obtain
\begin{eqnarray}
\label{doubleratios}
  &&  \ln \frac{\hat X_{2A}}{\hat X^2_{A}} +3 \ln A= 
  \ln\left[2^{7/2}\pi n_B \frac{a_\mathrm{a}^{1/2}}{T^2} \left(\frac{ \hbar^2}{m}\right)^{3/2} \frac{R_{2A}(T)}{R^2_A(T)}\right]\nonumber \\
  &&+(2-2^{2/3})\frac{a_\mathrm{s}}{T} A^{2/3}-0.714\,(2^{5/3}-2)Y_p^2  \frac{a_\mathrm{c}}{T}A^{5/3}.
\end{eqnarray}
The bulk term $a_\mathrm{b}$ and the asymmetry term $a_\mathrm{a}$ of the Bethe-Weizs{\"a}cker relation (\ref{BW}) cancel in the exponent.

\begin{figure*}[ht]
  \centering
    \includegraphics[width=\linewidth]{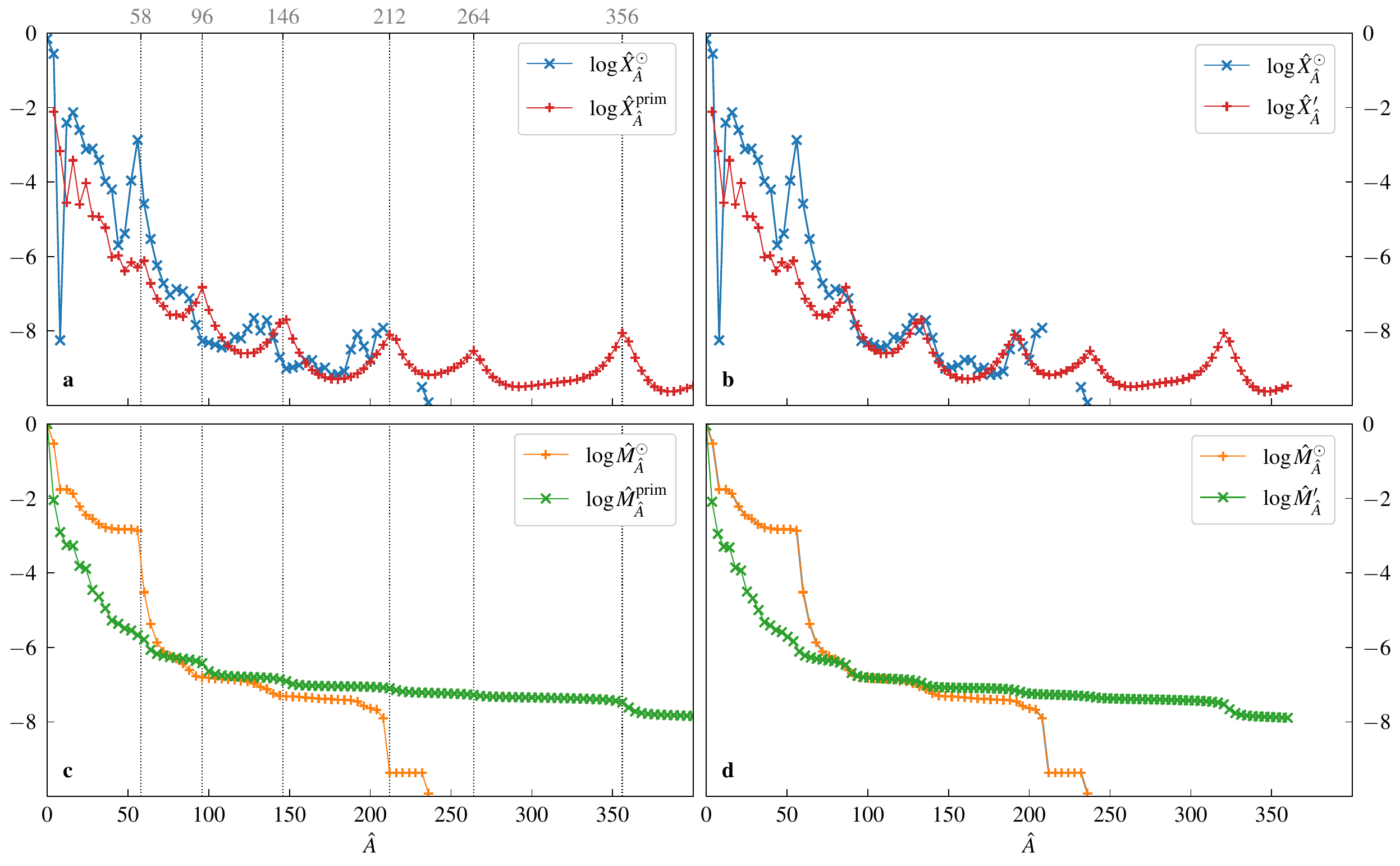}
   \caption{Accumulated mass fraction $\hat X_{\hat A}$ (top row) compared with the solar accumulated mass fraction $\hat X^\odot_{\hat A}$ (blue ``$\times$'' symbols) and $\hat A$-metallicity $M_{\hat A}$ (bottom row) compared with the solar $\hat A$-metallicity $M^\odot_{\hat A}$ (orange ``$+$'' symbols). Values are given in Table~\ref{tab.1} Dotted lines mark the double magic numbers $50\,{+}\,8$, $82\,{+}\,14$, $126\,{+}\,20$, $184\,{+}\,28$, $228\,{+}\,36$ and $308\,{+}\,50$ (included in $\hat A\,{=}\,356$).\\ 
   Panel \textbf{a}:
    Primordial accumulated mass fraction $\hat X^\mathrm{prim}_{\hat A}$ (red ``$+$'' symbols) for the parameter values $T=5.266\,\mathrm{MeV}$, $\mu_n=940.317\,\mathrm{MeV}$ and  $\mu_p=845.069\,\mathrm{MeV}$.\\
    Panel \textbf{b}:
    Accumulated mass fraction after evaporation of neutrons $\hat X'_{\hat A}$ (red ``$+$'' symbols) for $\bar{\nu} \,{=}\, 0.1 A$.\\
    Panel \textbf{c}:
   Primordial $\hat A$-metallicity $M_{\hat A}^\mathrm{prim}$ (green ``$\times$'' symbols)  for the parameter values $T=5.266\,\mathrm{MeV}$, $\mu_n=940.317\,\mathrm{MeV}$ and  $\mu_p=845.069\,\mathrm{MeV}$.\\
   Panel \textbf{d}:
   $\hat A$-metallicity $M_{\hat A}'$ after evaporation of neutrons (green ``$\times$'' symbols) for $\bar \nu =0.1 A$. \label{fig:XAshelln}}
\end{figure*}

We use this relation to infer values for $a_\mathrm{s}/T$ and $a_\mathrm{c}/T$ by comparison with the solar data $\hat X_{\hat A}^\odot$, Tab.~\ref{tab.1}, 
see Fig.~\ref{fig:ratios}. 
The strong deviations from a smooth behavior are  due to the fact that magic numbers are not included in the Bethe-Weizs{\"a}cker relation (\ref{BW}).
Relation (\ref{doubleratios}) is shown in Fig.~\ref{fig:ratios} for various parameter values $a_\mathrm{s}/T=1.6, \dots, 3.6$.
The graph is not very sensitive to the first term on the right-hand side of (\ref{doubleratios}) for which $a_\mathrm{a}\,{=}\,23.7\, \mathrm{MeV}$, $T\,{=}\,5\,\mathrm{MeV}$, and $n_\mathrm{B}\,{=}\,0.013\,\mathrm{fm}^{-3}$ has been used.
The least-mean-square-deviation fit to the data for $\hat A\,{=}\,(68,  72, 76, 80, 84, 88, 116)$, which are heavy nuclei not related to magic numbers, gives $a^\mathrm{matter}_\mathrm{s}/T\,{=}\,2.41$ and $Y_p^2 a^\mathrm{matter}_\mathrm{c}/T\,{=}\,0.00000244$. 
The small value of the Coulomb term in the LDM is an indication of a small value of the proton fraction $Y_p\approx 0.1$ and the strong screening in the dense, neutron-rich matter. 
Of interest is the result for the surface term $a_\mathrm{s}/T$.
Assuming a temperature of about $5 \, \mathrm{MeV}$, the value $a^\mathrm{matter}_\mathrm{s}\,{=}\,12.05\, \mathrm{MeV}$ follows, which is smaller than the free value $a_\mathrm{s}\,{=}\,17.81\,\mathrm{MeV}$ given above. The corresponding value $a_\mathrm{s}/T \approx 3.6$ is not compatible with the observed data, see Fig. \ref{fig:ratios}. 
A reduction of the surface term $a^\mathrm{matter}_\mathrm{s}$ for nuclei embedded in hot and dense matter is also discussed by 
\citet{1991NuPhA.535..331L}, see also \citet{Ravenhall:1983bdb}. 
The simple LDM discussed here supposes a value of the surface term in the range  $2 < a^\mathrm{matter}_\mathrm{s}/T<2.5$.

In a next step, we construct the primordial distribution. 
With the in-medium modifications of the LDM and the nucleon self-energy shifts given above, we determine the parameter values $T$, $\mu_n$ and $\mu_p$ so that $M^\mathrm{prim}_0\,{=}\,1$.
The assumption that the observed distribution $\hat X_{\hat A}$ of the heavy elements is  obtained from a primordial distribution predominately by decay processes (neutron emission, $\alpha$ decay, fission) has the consequence that the primordial $A$-metallicity $M^\mathrm{prim}_{\hat A}$ must have been larger than the presently observed $A$-metallicity $M^\odot_{\hat A}$ in the region $A \,{>}\, A_\mathrm{heavy} \,{\approx}\, 76$. 
At $\hat A\,{=}\,76$, only the evaporation of neutrons from the heavy nuclei significantly changes the primordial distribution $M_{\hat A}^{\rm prim}$ to the final one. 
The flow of matter bound in heavy nuclei across $A_\mathrm{heavy}$ due to $\alpha$ decay is small. It is determined by the number of nuclei which emit the $\alpha$ particles. Fission  gives no flow of matter across $A_\mathrm{heavy}$ because the fission fragments remain with mass numbers $A \,{>}\, A_\mathrm{heavy}$, i.e.\ the $A$-metallicity $M^\mathrm{prim}_{\hat A}$ in the range of $A_\mathrm{heavy}$ is conserved with respect to fission processes.
We consider these conditions for $M^\mathrm{prim}_{\hat A}$ in the region $76 \,{\le}\, \hat A \,{\le}\, 120$ as a prerequisite to infer the primordial distribution from the observed final abundances.

As an example, in Fig.~\ref{fig:XAshelln}, we show a result for parameter values $T\,{=}\, 5.266\, \mathrm{MeV}$,  $n_{\rm B}\,{=}\,0.013 \,\mathrm{fm}^{-3}$ and $Y_p\,{=}\,0.13$.
The chemical potentials are $\mu_n\,{=}\,940.317\, \mathrm{MeV}$ and  $\mu_p\,{=}\,845.069 \,\mathrm{MeV}$.
Here, the model parameter values  $a^\mathrm{matter}_{\rm b} \,{=}\,15.26\, \mathrm{MeV}$ and $a^\mathrm{matter}_{\rm s}=11 \,\mathrm{MeV}$ have been used. 
The parameter values for $T, \mu_n,\mu_p$ are obtained from the conditions that $M_0=1$ and $\Delta M_{\hat A}=M'_{\hat A}-M^\odot_{\hat A}\approx 0$ in the range $80\le \hat A \le 120$ as shown in Fig.~\ref{fig:XAshelln}, panel d.
The small deviation $\Delta M_{\hat A}>0$ is self-consistently determined by the $\alpha$ decay of heavy nuclei which is discussed below.

The inferred distribution extends to very large numbers of $A$, the accumulated primordial mass fraction $\hat X^\mathrm{prim}_{\hat A}$ drops down above $A \,{\approx}\, 600$.
The primordial $A$-metallicity $M^\mathrm{prim}_{\hat A}$ has also a long tail.
As shown in Fig.~\ref{fig:XAshelln} (c), it crosses the solar $M^\odot_{\hat A}$ near $\hat A=76$. 
The value of  $10^{-6.26}$ for $\hat A\,{=}\,76$ agrees well with the observed values, implying that the gain of heavy nuclei matter by fission of light nuclei and neutron absorption is compensated by the loss of heavy nuclei matter due to evaporation of neutrons and $\alpha$ decay.
We consider $\hat A\,{=}\,76$ as the border between light and heavy nuclei. The accumulated solar mass fraction takes a minimum here. 
Burning processes will hardly go much behind the iron peak, and fission of superheavy elements and emission of clusters are also not expected to contribute to a flow across this value. 
Emission of neutrons is possible but will give only a small left-directed flow, which is compensated by the right-directed fusion flow.
The metallicity $M_{76}$ can approximately be considered  as conserved quantity. We expect only minor changes of its value after HEFO. Fusion will change the distribution of the light nuclei after HEFO, forming, for example, the iron peak.
We do not discuss the light nuclei $A \,{<}\, 76$ here.

For the heavy nuclei, we obtain three peaks at mass numbers near the observed maxima, but at slightly higher values of $A$, as shown in Fig.~\ref{fig:XAshelln} (a).
These and further maxima are related to magic numbers of neutrons and protons,  $58\,{=}\,50+8$, $96\,{=}\,82+14$, $146\,{=}\,126+20$ and $212\,{=}\,184+28$. The proton fractions for these double-magic nuclei are close to the assumed proton fraction $Y_p=0.13$ of the nuclear matter.
 The corresponding nuclei are bound but far from stability. The neutron separation energy $S_n$ is negative, but they nonetheless contribute as correlations in the continuum  to the composition of matter.
For example, as discussed above in Sect.~\ref{Labor}, it is shown by experiments that the unstable nucleus $^5$He ($S_n < 0$)  contributes to the primordial distribution in ternary fission. 
The appearance of peaks at higher mass numbers $\hat A \,{=}\, 264$ and $356$ is of interest.

De-excitation processes will modify the primordial distribution. $\beta$- and $\gamma$-decays do not change the distribution of mass fractions.
The evaporation of neutrons concerns all heavy nuclei which are excited. 
Using Eq.~(\ref{evap}) with $d \,T^2_{\rm MeV}\,{=}\,0.1$, the evaporation of neutrons results in the mass fraction distribution $X_{A}'$ shown in panel b of Fig.~\ref{fig:XAshelln}. A corresponding plot of the accumulated mass fractions $\hat X'_{\hat A}$ shows a good reproduction of the three peaks.

The accumulated mass fraction $\hat X'_{\hat A}$  extends significantly beyond the limit of stability at $A\,{=}\,208$.
Nuclei with $A\,{>}\,208$ are unstable with respect to $\alpha$-decay, binary fission, ternary fission and multifragmentation.
$\alpha$-decays  shift the actinides  toward the lead region. 
The overpopulation of accumulated mass fractions $\hat X^\odot_{204}$ and $\hat X^\odot_{208}$ accounts for part of the actinides. 
For an estimate, we calculate mass fraction of the overpopulation $\hat X^\odot_{204}+\hat X^\odot_{208}-\hat X'_{204}+\hat X'_{208}=1.92 \times 10^{-8}$.
The amount of matter bound in nuclei with mass numbers $A\ge212$ to decay via fission or via $\alpha$ decay is $M'_{212}=5.44 \times 10^{-8}$.
For the estimate, let us make the assumption of a sharp separation between nuclei with smaller mass numbers $A$, which decay by emission of an $\alpha$ particle, and superheavy nuclei with larger mass numbers, which decay by fission.
 Of course, the branching ratios for fission and $\alpha$ decay of superheavy nuclei is not sharply jumping with $A$.
In this simple estimate, since $M'_{312}=3.52 \times 10^{-8}$, the fraction of matter bound in nuclei with $212 \le A \le 312$ is required to account for the overpopulation at $\hat A= 204, 208$. 
The loss of matter bound in heavy nuclei due to $\alpha$-decay is estimated as $\Delta X_{\alpha}=\sum \hat X'_{\hat A} (1-208/\hat A) \approx 2\times 10^{-9}$ over this interval.
This determines the difference $\Delta M_{\hat A}$ given above in the interval $80 \le \hat A \le 120$.
The process of $\alpha$-decay is not completed yet; reduced values of the accumulated mass fractions of $\hat A \,{=}\,232$ and $236$ (Th, U, Pu) remain and are subject to chemochronology.

Fission is the fate of the remaining heavy nuclei, with a total mass fraction of $3.52 \times 10^{-8}$ according to our estimate.
Of  particular relevance is the peak at $A\,{=}\,356$, which is related to the double magic nucleus $^{356}$Sn.
After evaporation of neutrons, the maximum appears in the distribution $\hat X'_{\hat A}$ near $\hat A \,{=}\,320$. 
Assuming symmetric fission \citep{PhysRevC.103.025806}, fragments near $A\,{=}\,106$ would appear, which can explain the enhanced observed mass fractions in that region.

The example of the freeze-out distribution considered here is determined by the Lagrange parameters $T,\mu_n,\mu_p$ which yield the primordial $A$-metallicity $M^\mathrm{prim}_{\hat A}$ shown in Panel c of Fig.~\ref{fig:XAshelln}, see also Tab.~\ref{tab.1}. The inferred parameter values $T\,{=}\, 5.266\, \mathrm{MeV}$, 
 $\mu_n\,{=}\,940.317\, \mathrm{MeV}$ and  $\mu_p\,{=}\,845.069 \,\mathrm{MeV}$ fulfill the requirements  $M_0=1$ and $M_{\hat A}^{\rm prim} > M^\odot_{\hat A}$ for $\hat A> A_{\rm heavy}$ given at the beginning of this section. 
After evaporation of neutrons, $M'_{\hat A}$ coincides with $M^\odot_{\hat A}$ in the range $100 \,{\le}\, \hat A \,{\le}\, 120$ with the small gap $\Delta M_{\alpha}$ which accounts for the $\alpha$-decay. 
Taking into account the $\alpha$ decay and fission, the $\hat A$ metallicity $M'_{\hat A}$ will be further changed so that it approaches $M^\odot_{\hat A}$.
We investigated only the gross structure of the final distribution of mass fractions $\hat X_{\hat A}$ which concerns the sum of the detailed mass fractions $X_{A,Z}$ with four subsequent mass numbers $A$. In order to obtain the final distribution of mass fractions $X_{A,Z}$, a more detailed description of the individual properties of nuclei is necessary, for instance of the decay processes of the various excited isotopes far of the region of stability.

\section{Discussion and conclusions}
\label{sec:discussion}

Within the framework of a non-equilibrium approach, which includes both the kinetic and the hydrodynamic approximation for the description of expanding dense matter,
parameter values for the freeze-out of heavy element formation ($A \,{\ge}\, 76$) are inferred. 
Under our specific approximations, HEFO parameter values $T\,{=}\,5.266\,\mathrm{MeV}$, $n_\mathrm{B}\,{=}\,0.013\, \mathrm{fm}^{-3}$, and $Y_p\,{=}\,0.13$ are estimated, which characterise the astrophysical site of formation of the heavy elements with the characteristic solar system abundance pattern. 
Regardless of the specific astrophysical event (supernova explosion, NS merger or inhomogeneous Big Bang), these relevant parameter values are an important prerequisite for discussing the origin of the observed abundances. 

A hydrodynamic approach based on the assumption of local thermodynamic equilibrium is more suitable if the relaxation time to equilibrium is sufficiently short. Although
NRN simulations can describe situations with short relaxation times, so that the system reaches a quasi-equilibrium, they fail to reproduce the correct equilibrium state, which also contains in-medium corrections and continuum correlations\footnote{In analogy, the Boltzmann equation has the ideal gas as the equilibrium solution; conservation of energy, however, requires the inclusion of higher-order correlation functions.}. 

We therefore take the alternative approach of a local thermodynamic equilibrium description of matter. In this framework, the relevant statistical operator accounts for all equilibrium correlations. A similar situation is encountered for HIC at the highest energies \citep{Andronic:2017pug}, for which we know that correlations in the continuum (up to $2 \, \mathrm{GeV}$ excitation energy), as described by the hadron resonance gas, are relevant for the observed yields. Our freeze-out scenario describes expanding dense matter in which relaxation to the generalized Gibbs state (\ref{Gibbseq}) occurs as long as the relaxation time for reactive collisions is sufficiently short.
The $r$-process reactions, which lead to increasing mass numbers of heavy nuclei, require high neutron densities in the environment.
As the neutron density decreases, the reaction rates also become smaller. Eventually, at the HEFO, the distribution of the heavy elements freezes out.
As  mentioned by \citet{1987PhLB..185..281R}, the HEFO parameter values lie in the thermodynamic instability region of the nuclear matter (liquid--gas) phase transition, see also \citet{2003NuPhA.723..517I}, so that a rapid change in the parameter values -- especially that of the baryon density -- is to be expected during expansion.
The connection of freeze-out and the formation of bound states (Mott effect) in nuclear matter has recently been discussed by \citet{blaschke2024a}.
The investigation of the dynamical evolution of hot and dense matter under these conditions is beyond the scope of this work and will be considered in the future.

After HEFO, the excited, unstable states of nuclei which exist in the primordial distribution will decay and feed the abundances of nuclei with  lower mass numbers $A$.
The NRN currently used to simulate the $r$-process 
are well established to describe chemical evolution in the low-density regime ($n_\mathrm{B} \,{\le}\, 10^{-4}$ fm$^{-3}$) where in-medium effects are less important. 
For subsaturation densities ($0.0001 \,\mathrm{fm}^{-3} \,{\le}\, n_\mathrm{B} \,{\le}\, 0.15$ fm$^{-3}$), however, Pauli blocking and the Mott effect must be taken into account. The concept of NSE \citep{Yudin:2018ckx} cannot be applied in this regime and leads to flawed results as shown by \citet{Fischer:2020krf}.

Applying our method to an astrophysical situation, we argue that the relevant statistical operator at freeze-out (\ref{Gibbseq}) provides an appropriate tool to determine the initial state for performing NRN calculations describing the chemical evolution during the time after freeze-out. 

As a consequence of our freeze-out scenario, the assumed path of the $r$-process near the neutron-drip line is no longer relevant. 
Instead, the matter in the high-density site of nucleosynthesis is described to be in the generalized Gibbs state (\ref{Gibbseq}), where highly excited nuclei and continuum correlations are present before nucleosynthesis of heavy elements starts.

It is also not expected that the formation of heavy elements would terminate in the neutron-rich superheavy region due to fission and rarely go beyond the element $Z=100$ \citep{Chen_2023}.
Instead, the trajectory of matter in the density-temperature plane passes through states characterized by the Lagrangian parameters given in this work (see Fig.~\ref{fig:varsize}). 
Nuclei with mass numbers as large as $A \,{\approx}\, 600$ are formed in the primordial distribution. Fission and $\alpha$ decay are a common side effects of nucleosynthesis. Hence, an interesting prediction of this scenario is the increased primordial abundance of nuclei in the range around $A\,{=}\,358, Z\,{=}\,50$, which due to fission contribute significantly to the observed increased abundances in the range $A\,{=}\,160$.

Our proposed scenario raises the question of the astrophysical sites of heavy-element formation. We will explore in future studies whether SN explosions and NS mergers reach the conditions required for the production of heavy elements outlined in this work. At present, it remains unclear to what extent they contribute to the observed abundances. 
We therefore point out another possibility -- the inhomogeneous Big-Bang nucleosynthesis (IBBN) scenario, in which matter has already clumped into high-density objects before Big-Bang nucleosynthesis commences. 
Most of the matter in this scenario never went through the composition $X_{A,Z}^\mathrm{HBBR}$ as given in Eq.~(\ref{hBBN}). This may explain the so-far unsuccessful search for population III stars and pair-instability supernovae. Moreover, the IBBN scenario coupled to our freeze-out model for the formation of heavy elements offers 
a natural explanation of the universal abundance pattern
beyond the iron group found in various astrophysical objects as well as in the solar system. In this framework,
only the three parameter values for $T$ and  $\mu_\tau$  determine the full primordial distribution, from which the observed abundance pattern freezes out according to nuclear properties. Given the recent JWST observations of massive objects at large redshifts \citep{2023ApJ...947L..24M} and the reports of signaturs of unusual nucleosynthesis from a massive star in the early Universe \citep{Ji:2024edf}, an IBBN scenario may not be unrealistic.

In conclusion, we show that the gross structure of the solar mass fraction distribution beyond $A \,{=}\,76$ can be explained in terms of a freeze-out process from a universal primordial equilibrium state characterized by a minimal set of parameters. Astrophysical sites where suitable parameters are encountered will be explored in follow-up studies.

\begin{acknowledgements}
GR acknowledges a stipend from the Foundation for Polish Science within the Alexander von Humboldt programme under grant No. DPN/JJL/402-4773/2022. 
DB was supported by NCN under grant No. 2021/43/P/ST2/03319. 
The work of FKR  is supported by the Klaus Tschira Foundation, by the Deutsche Forschungsgemeinschaft (DFG, German Research Foundation) -- RO 3676/7-1, project number 537700965,
and by the European Union (ERC, ExCEED, project number 101096243). Views and opinions expressed are, however, those of the authors only and do not necessarily reflect those of the European Union or the European Research Council Executive Agency. Neither the European Union nor the granting authority can be held responsible for them.
\end{acknowledgements}


\begin{thebibliography}{103}
\expandafter\ifx\csname natexlab\endcsname\relax\def\natexlab#1{#1}\fi

\bibitem[{Andronic {et~al.}(2018)Andronic, Braun-Munzinger, Redlich, \&
  Stachel}]{Andronic:2017pug}
Andronic, A., Braun-Munzinger, P., Redlich, K., \& Stachel, J. 2018, Nature,
  561, 321

\bibitem[{Arcones \& Thielemann(2013)}]{Arcones:2012wj}
Arcones, A. \& Thielemann, F.~K. 2013, J. Phys. G, 40, 013201

\bibitem[{Arcones \& Thielemann(2023)}]{Arcones:2022jer}
Arcones, A. \& Thielemann, F.-K. 2023, Astron. Astrophys. Rev., 31, 1

\bibitem[{{Asplund} {et~al.}(2009){Asplund}, {Grevesse}, {Sauval}, \&
  {Scott}}]{Asplund09}
{Asplund}, M., {Grevesse}, N., {Sauval}, A.~J., \& {Scott}, P. 2009, \araa, 47,
  481

\bibitem[{{Beun} {et~al.}(2008){Beun}, {McLaughlin}, {Surman}, \&
  {Hix}}]{2008PhRvC..77c5804B}
{Beun}, J., {McLaughlin}, G.~C., {Surman}, R., \& {Hix}, W.~R. 2008, \prc, 77,
  035804

\bibitem[{{Blanchard} {et~al.}(2024){Blanchard}, {Villar}, {Chornock},
  {Laskar}, {Li}, {Leja}, {Pierel}, {Berger}, {Margutti}, {Alexander},
  {Barnes}, {Cendes}, {Eftekhari}, {Kasen}, {LeBaron}, {Metzger}, {Muzerolle
  Page}, {Rest}, {Sears}, {Siegel}, \& {Yadavalli}}]{2024NatAs...8..774B}
{Blanchard}, P.~K., {Villar}, V.~A., {Chornock}, R., {et~al.} 2024, Nature
  Astronomy, 8, 774

\bibitem[{{Blaschke} {et~al.}(2024){Blaschke}, {Liebing}, {R{\"o}pke}, \&
  {D{\"o}nigus}}]{blaschke2024a}
{Blaschke}, D., {Liebing}, S., {R{\"o}pke}, G., \& {D{\"o}nigus}, B. 2024,
  arXiv e-prints, arXiv:2408.01399

\bibitem[{{Bohr} \& {Mottelson}(1969)}]{BohrM}
{Bohr}, A. \& {Mottelson}, B. 1969, Nuclear Structure (W.A. Benjamin, Inc., New
  York)

\bibitem[{Burbidge {et~al.}(1957)Burbidge, Burbidge, Fowler, \&
  Hoyle}]{Burbidge:1957vc}
Burbidge, M.~E., Burbidge, G.~R., Fowler, W.~A., \& Hoyle, F. 1957, Rev. Mod.
  Phys., 29, 547

\bibitem[{Cameron(1957)}]{Cameron_1957}
Cameron, A. G.~W. 1957, Publications of the Astronomical Society of the
  Pacific, 69, 201

\bibitem[{{Chen} {et~al.}(2024){Chen}, {Landry}, {Read}, \&
  {Siegel}}]{Chen:2024gwj}
{Chen}, H.-Y., {Landry}, P., {Read}, J.~S., \& {Siegel}, D.~M. 2024, arXiv
  e-prints, arXiv:2402.03696

\bibitem[{Chen {et~al.}(2023)Chen, Pei, Qiang, \& Chi}]{Chen_2023}
Chen, J., Pei, J., Qiang, Y., \& Chi, J. 2023, Chinese Physics Letters, 40,
  012401

\bibitem[{Chen {et~al.}(2024)Chen, Li, Chen, Hu, \& Liang}]{Chen:2024acv}
Chen, M.-H., Li, L.-X., Chen, Q.-H., Hu, R.-C., \& Liang, E.-W. 2024, Mon. Not.
  Roy. Astron. Soc., 529, 1154

\bibitem[{{Coc} \& {Vangioni}(2017)}]{2017IJMPE..2641002C}
{Coc}, A. \& {Vangioni}, E. 2017, International Journal of Modern Physics E,
  26, 1741002

\bibitem[{{Cowan} {et~al.}(1995){Cowan}, {Burris}, {Sneden}, {McWilliam}, \&
  {Preston}}]{cowan1995a}
{Cowan}, J.~J., {Burris}, D.~L., {Sneden}, C., {McWilliam}, A., \& {Preston},
  G.~W. 1995, \apjl, 439, L51

\bibitem[{Cowan {et~al.}(2021)Cowan, Sneden, Lawler, Aprahamian, Wiescher,
  Langanke, Mart\'\i{}nez-Pinedo, \& Thielemann}]{Cowan:2019pkx}
Cowan, J.~J., Sneden, C., Lawler, J.~E., {et~al.} 2021, Rev. Mod. Phys., 93,
  15002

\bibitem[{{Cyburt} {et~al.}(2016){Cyburt}, {Fields}, {Olive}, \&
  {Yeh}}]{2016RvMP...88a5004C}
{Cyburt}, R.~H., {Fields}, B.~D., {Olive}, K.~A., \& {Yeh}, T.-H. 2016, Reviews
  of Modern Physics, 88, 015004

\bibitem[{{Dieperink} \& {van Isacker}(2009)}]{2009EPJA...42..269D}
{Dieperink}, A.~E.~L. \& {van Isacker}, P. 2009, European Physical Journal A,
  42, 269

\bibitem[{{Dinh Thi} {et~al.}(2023){Dinh Thi}, {Fantina}, \&
  {Gulminelli}}]{2023EPJA...59..292D}
{Dinh Thi}, H., {Fantina}, A.~F., \& {Gulminelli}, F. 2023, European Physical
  Journal A, 59, 292

\bibitem[{Dinh~Thi {et~al.}(2023)Dinh~Thi, Fantina, \&
  Gulminelli}]{DinhThi:2023ioy}
Dinh~Thi, H., Fantina, A.~F., \& Gulminelli, F. 2023, Astron. Astrophys., 672,
  A160

\bibitem[{{Duflo} \& {Zuker}(1995)}]{1995PhRvC..52...23D}
{Duflo}, J. \& {Zuker}, A.~P. 1995, \prc, 52, R23

\bibitem[{Eichler {et~al.}(2015)}]{Eichler:2014kma}
Eichler, M. {et~al.} 2015, Astrophys. J., 808, 30

\bibitem[{{Ernandes} {et~al.}(2023){Ernandes}, {Castro}, {Barbuy}, {Spite},
  {Hill}, {Castilho}, \& {Evans}}]{ernandez2023a}
{Ernandes}, H., {Castro}, M.~J., {Barbuy}, B., {et~al.} 2023, \mnras, 524, 656

\bibitem[{{Fetter} \& {Walecka}(1971)}]{1971qtmp.book.....F}
{Fetter}, A.~L. \& {Walecka}, J.~D. 1971, {Quantum theory of many-particle
  systems} (San Francisco, McGraw-Hill)

\bibitem[{Fields {et~al.}(2014)Fields, Molaro, \& Sarkar}]{Fields:2014uja}
Fields, B.~D., Molaro, P., \& Sarkar, S. 2014, Chin. Phys. C, 38, 339

\bibitem[{Fischer {et~al.}(2024)Fischer, Guo, Langanke, Martinez-Pinedo, Qian,
  \& Wu}]{Fischer:2023ebq}
Fischer, T., Guo, G., Langanke, K., {et~al.} 2024, Prog. Part. Nucl. Phys.,
  137, 104107

\bibitem[{{Fischer} {et~al.}(2014){Fischer}, {Hempel}, {Sagert}, {Suwa}, \&
  {Schaffner-Bielich}}]{2014EPJA...50...46F}
{Fischer}, T., {Hempel}, M., {Sagert}, I., {Suwa}, Y., \& {Schaffner-Bielich},
  J. 2014, European Physical Journal A, 50, 46

\bibitem[{Fischer {et~al.}(2020)Fischer, Typel, R\"opke, Bastian, \&
  Mart\'\i{}nez-Pinedo}]{Fischer:2020krf}
Fischer, T., Typel, S., R\"opke, G., Bastian, N.-U.~F., \&
  Mart\'\i{}nez-Pinedo, G. 2020, Phys. Rev. C, 102, 055807

\bibitem[{Fraisse {et~al.}(2023)Fraisse, B\'elier, M\'eot, Gaudefroy,
  Francheteau, \& Roig}]{Fraisse:2023hfx}
Fraisse, B., B\'elier, G., M\'eot, V., {et~al.} 2023, Phys. Rev. C, 108, 014610

\bibitem[{{Frebel}(2018)}]{frebel2018a}
{Frebel}, A. 2018, Annual Review of Nuclear and Particle Science, 68, 237

\bibitem[{{Frebel} \& {Norris}(2015)}]{2015ARA&A..53..631F}
{Frebel}, A. \& {Norris}, J.~E. 2015, \araa, 53, 631

\bibitem[{Furusawa \& Nagakura(2023)}]{Furusawa:2022ktu}
Furusawa, S. \& Nagakura, H. 2023, Prog. Part. Nucl. Phys., 129, 104018

\bibitem[{Furusawa {et~al.}(2017)Furusawa, Togashi, Nagakura, Sumiyoshi,
  Yamada, Suzuki, \& Takano}]{Furusawa:2017auz}
Furusawa, S., Togashi, H., Nagakura, H., {et~al.} 2017, J. Phys. G, 44, 094001

\bibitem[{{Goriely} {et~al.}(2009){Goriely}, {Chamel}, \&
  {Pearson}}]{2009PhRvL.102o2503G}
{Goriely}, S., {Chamel}, N., \& {Pearson}, J.~M. 2009, \prl, 102, 152503

\bibitem[{Goriely \& Kullmann(2020)}]{Goriely2020}
Goriely, S. \& Kullmann, I. 2020, R-Process Nucleosynthesis in Neutron Star
  Merger Ejecta and Nuclear Dependences, ed. I.~Tanihata, H.~Toki, \& T.~Kajino
  (Singapore: Springer Nature Singapore), 1--26

\bibitem[{Gulminelli \& Raduta(2015)}]{Gulminelli:2015csa}
Gulminelli, F. \& Raduta, A.~R. 2015, Phys. Rev. C, 92, 055803

\bibitem[{{Heger} \& {Woosley}(2002)}]{heger2002a}
{Heger}, A. \& {Woosley}, S.~E. 2002, \apj, 567, 532

\bibitem[{Hempel \& Schaffner-Bielich(2010)}]{Hempel:2009mc}
Hempel, M. \& Schaffner-Bielich, J. 2010, Nucl. Phys. A, 837, 210

\bibitem[{{Hill} {et~al.}(2002){Hill}, {Plez}, {Cayrel}, {Beers},
  {Nordstr{\"o}m}, {Andersen}, {Spite}, {Spite}, {Barbuy}, {Bonifacio},
  {Depagne}, {Fran{\c{c}}ois}, \& {Primas}}]{hill2002a}
{Hill}, V., {Plez}, B., {Cayrel}, R., {et~al.} 2002, \aap, 387, 560

\bibitem[{{Hillebrandt} {et~al.}(1984){Hillebrandt}, {Nomoto}, \&
  {Wolff}}]{1984A&A...133..175H}
{Hillebrandt}, W., {Nomoto}, K., \& {Wolff}, R.~G. 1984, \aap, 133, 175

\bibitem[{{Iliadis}(2015)}]{iliadis2015a}
{Iliadis}, C. 2015, Nuclear physics of stars (Wiley-VCH, Weinheim)

\bibitem[{{Ishizuka} {et~al.}(2003){Ishizuka}, {Ohnishi}, \&
  {Sumiyoshi}}]{2003NuPhA.723..517I}
{Ishizuka}, C., {Ohnishi}, A., \& {Sumiyoshi}, K. 2003, \nphysa, 723, 517

\bibitem[{Janka \& Bauswein(2020)}]{Janka2020}
Janka, H.-T. \& Bauswein, A. 2020, Dynamics and Equation of State Dependencies
  of Relevance for Nucleosynthesis in Supernovae and Neutron Star Mergers, ed.
  I.~Tanihata, H.~Toki, \& T.~Kajino (Singapore: Springer Nature Singapore),
  1--98

\bibitem[{Ji {et~al.}(2024)}]{Ji:2024edf}
Ji, A.~P. {et~al.} 2024, Astrophys. J. Lett., 961, L41

\bibitem[{{Just} {et~al.}(2023){Just}, {Vijayan}, {Xiong}, {Goriely},
  {Soultanis}, {Bauswein}, {Guilet}, {Janka}, \&
  {Mart{\'\i}nez-Pinedo}}]{2023ApJ...951L..12J}
{Just}, O., {Vijayan}, V., {Xiong}, Z., {et~al.} 2023, \apjl, 951, L12

\bibitem[{{Kobayashi} {et~al.}(2020){Kobayashi}, {Karakas}, \&
  {Lugaro}}]{2020ApJ...900..179K}
{Kobayashi}, C., {Karakas}, A.~I., \& {Lugaro}, M. 2020, \apj, 900, 179

\bibitem[{Kopatch {et~al.}(2002)Kopatch, Mutterer, Schwalm, Thirolf, \&
  Gonnenwein}]{Kopatch:2002bd}
Kopatch, Y.~N., Mutterer, M., Schwalm, D., Thirolf, P., \& Gonnenwein, F. 2002,
  Phys. Rev. C, 65, 044614

\bibitem[{K\"oster {et~al.}(1999)K\"oster, Faust, Fioni, Friedrichs, Gro\ss{},
  \& Oberstedt}]{Koster:1999hpd}
K\"oster, U., Faust, H., Fioni, G., {et~al.} 1999, Nucl. Phys. A, 652, 371

\bibitem[{{Koura} \& {Chiba}(2013)}]{2013JPSJ...82a4201K}
{Koura}, H. \& {Chiba}, S. 2013, Journal of the Physical Society of Japan, 82,
  014201

\bibitem[{{Kraeft} {et~al.}(1986){Kraeft}, {Kremp}, {Ebeling}, \&
  {R{\"o}pke}}]{KKER}
{Kraeft}, W.-D., {Kremp}, D., {Ebeling}, W., \& {R{\"o}pke}, G. 1986, Quantum
  Statistics of Charged Particle Systems (Akademie-Verlag, Berlin)

\bibitem[{{Lattimer} \& {Swesty}(1991)}]{1991NuPhA.535..331L}
{Lattimer}, J.~M. \& {Swesty}, D.~F. 1991, \nphysa, 535, 331

\bibitem[{Lema\^{\i}tre {et~al.}(2021)Lema\^{\i}tre, Goriely, Bauswein, \&
  Janka}]{PhysRevC.103.025806}
Lema\^{\i}tre, J.-F., Goriely, S., Bauswein, A., \& Janka, H.-T. 2021, Phys.
  Rev. C, 103, 025806

\bibitem[{Lippuner \& Roberts(2017)}]{Lippuner:2017tyn}
Lippuner, J. \& Roberts, L.~F. 2017, Astrophys. J. Suppl., 233, 18

\bibitem[{{Lodders}(2003)}]{Lodders03}
{Lodders}, K. 2003, \apj, 591, 1220

\bibitem[{{Lodders}(2021)}]{2021SSRv..217...44L}
{Lodders}, K. 2021, \ssr, 217, 44

\bibitem[{{Lodders} {et~al.}(2009){Lodders}, {Palme}, \&
  {Gail}}]{2009LanB...4B..712L}
{Lodders}, K., {Palme}, H., \& {Gail}, H.~P. 2009, Landolt B{\"o}rnstein, 4B,
  712

\bibitem[{{Magg} {et~al.}(2022){Magg}, {Bergemann}, {Serenelli}, {Bautista},
  {Plez}, {Heiter}, {Gerber}, {Ludwig}, {Basu}, {Ferguson}, {Gallego},
  {Gamrath}, {Palmeri}, \& {Quinet}}]{magg2022a}
{Magg}, E., {Bergemann}, M., {Serenelli}, A., {et~al.} 2022, \aap, 661, A140

\bibitem[{{M{\"o}ller} {et~al.}(2016){M{\"o}ller}, {Sierk}, {Ichikawa}, \&
  {Sagawa}}]{2016ADNDT.109....1M}
{M{\"o}ller}, P., {Sierk}, A.~J., {Ichikawa}, T., \& {Sagawa}, H. 2016, Atomic
  Data and Nuclear Data Tables, 109, 1

\bibitem[{{Morishita} {et~al.}(2023){Morishita}, {Roberts-Borsani}, {Treu},
  {Brammer}, {Mason}, {Trenti}, {Vulcani}, {Wang}, {Acebron}, {Bah{\'e}},
  {Bergamini}, {Boyett}, {Bradac}, {Calabr{\`o}}, {Castellano}, {Chen}, {De
  Lucia}, {Filippenko}, {Fontana}, {Glazebrook}, {Grillo}, {Henry}, {Jones},
  {Kelly}, {Koekemoer}, {Leethochawalit}, {Lu}, {Marchesini}, {Mascia},
  {Mercurio}, {Merlin}, {Metha}, {Nanayakkara}, {Nonino}, {Paris},
  {Pentericci}, {Rosati}, {Santini}, {Strait}, {Vanzella}, {Windhorst}, \&
  {Xie}}]{2023ApJ...947L..24M}
{Morishita}, T., {Roberts-Borsani}, G., {Treu}, T., {et~al.} 2023, \apjl, 947,
  L24

\bibitem[{{Munoz} {et~al.}(2024){Munoz}, {Udrescu}, \& {Garcia
  Ruiz}}]{2024arXiv240411477M}
{Munoz}, J.~M., {Udrescu}, S.~M., \& {Garcia Ruiz}, R.~F. 2024, arXiv e-prints,
  arXiv:2404.11477

\bibitem[{{Nakamura} {et~al.}(2017){Nakamura}, {Hashimoto}, {Ichimasa}, \&
  {Arai}}]{2017IJMPE..2641003N}
{Nakamura}, R., {Hashimoto}, M.-A., {Ichimasa}, R., \& {Arai}, K. 2017,
  International Journal of Modern Physics E, 26, 1741003

\bibitem[{{National Nuclear Data Center}(2024)}]{nuclei}
{National Nuclear Data Center}. 2024, {NuDat 3.0},
  \url{https://www.nndc.bnl.gov/nudat/}, accessed: 2024-07-28

\bibitem[{{National Research Council}(2003)}]{NAP10079}
{National Research Council}. 2003, Connecting Quarks with the Cosmos: Eleven
  Science Questions for the New Century (Washington, DC: The National Academies
  Press)

\bibitem[{Natowitz {et~al.}(2023)Natowitz, Pais, \& R\"opke}]{Natowitz:2022npi}
Natowitz, J.~B., Pais, H., \& R\"opke, G. 2023, Phys. Rev. C, 107, 014618

\bibitem[{{Natowitz} {et~al.}(2010){Natowitz}, {R{\"o}pke}, {Typel},
  {Blaschke}, {Bonasera}, {Hagel}, {Kl{\"a}hn}, {Kowalski}, {Qin}, {Shlomo},
  {Wada}, \& {Wolter}}]{Natowitz10}
{Natowitz}, J.~B., {R{\"o}pke}, G., {Typel}, S., {et~al.} 2010, \prl, 104,
  202501

\bibitem[{Natowitz {et~al.}(2010)}]{Natowitz:2010ti}
Natowitz, J.~B. {et~al.} 2010, Phys. Rev. Lett., 104, 202501

\bibitem[{Pais {et~al.}(2015)Pais, Chiacchiera, \&
  Provid\^encia}]{Pais:2015xoa}
Pais, H., Chiacchiera, S., \& Provid\^encia, C. 2015, Phys. Rev. C, 91, 055801

\bibitem[{Pais {et~al.}(2019)Pais, Gulminelli, Provid\^encia, \&
  R\"opke}]{Pais:2019shp}
Pais, H., Gulminelli, F., Provid\^encia, C., \& R\"opke, G. 2019, Phys. Rev. C,
  99, 055806

\bibitem[{Panov(2023)}]{Panov:2023tfn}
Panov, I.~V. 2023, Phys. Part. Nucl., 54, 542

\bibitem[{{Panov} {et~al.}(2018){Panov}, {Glazyrin}, {R{\"o}pke}, \&
  {Blinnikov}}]{panov2018a}
{Panov}, I.~V., {Glazyrin}, S.~I., {R{\"o}pke}, F.~K., \& {Blinnikov}, S.~I.
  2018, Astronomy Letters, 44, 309

\bibitem[{Pochodzalla {et~al.}(1995)}]{Pochodzalla:1995xy}
Pochodzalla, J. {et~al.} 1995, Phys. Rev. Lett., 75, 1040

\bibitem[{Qin {et~al.}(2012)}]{Qin:2011qp}
Qin, L. {et~al.} 2012, Phys. Rev. Lett., 108, 172701

\bibitem[{Rauscher(2003)}]{Rauscher:2003ti}
Rauscher, T. 2003, Astrophys. J. Suppl., 147, 403

\bibitem[{Ravenhall {et~al.}(1983)Ravenhall, Pethick, \&
  Lattimer}]{Ravenhall:1983bdb}
Ravenhall, D.~G., Pethick, C.~J., \& Lattimer, J.~M. 1983, Nucl. Phys. A, 407,
  571

\bibitem[{{Reichert} {et~al.}(2023){Reichert}, {Winteler}, {Korobkin},
  {Arcones}, {Bliss}, {Eichler}, {Frischknecht}, {Fr{\"o}hlich}, {Hirschi},
  {Jacobi}, {Kuske}, {Mart{\'\i}nez-Pinedo}, {Martin}, {Mocelj}, {Rauscher}, \&
  {Thielemann}}]{reichert2023a}
{Reichert}, M., {Winteler}, C., {Korobkin}, O., {et~al.} 2023, \apjs, 268, 66

\bibitem[{{Ricigliano} {et~al.}(2024){Ricigliano}, {Jacobi}, \&
  {Arcones}}]{ricigliano2024}
{Ricigliano}, G., {Jacobi}, M., \& {Arcones}, A. 2024, \mnras, 533, 2096

\bibitem[{{Roederer} {et~al.}(2009){Roederer}, {Kratz}, {Frebel}, {Christlieb},
  {Pfeiffer}, {Cowan}, \& {Sneden}}]{roederer2009a}
{Roederer}, I.~U., {Kratz}, K.-L., {Frebel}, A., {et~al.} 2009, \apj, 698, 1963

\bibitem[{{Roederer} {et~al.}(2022){Roederer}, {Lawler}, {Den Hartog},
  {Placco}, {Surman}, {Beers}, {Ezzeddine}, {Frebel}, {Hansen}, {Hattori},
  {Holmbeck}, \& {Sakari}}]{2022ApJS..260...27R}
{Roederer}, I.~U., {Lawler}, J.~E., {Den Hartog}, E.~A., {et~al.} 2022, \apjs,
  260, 27

\bibitem[{Roederer {et~al.}(2023)}]{Roederer:2023spd}
Roederer, I.~U. {et~al.} 2023, Science, 382, adf1341

\bibitem[{Rohlf(1994)}]{rohlf1994a}
Rohlf, J.~W. 1994, Modern Physics from [alpha] to Z° (Wiley)

\bibitem[{{R{\"o}pke}(1987)}]{1987PhLB..185..281R}
{R{\"o}pke}, G. 1987, Physics Letters B, 185, 281

\bibitem[{{R{\"o}pke}(2009)}]{2009PhRvC..79a4002R}
{R{\"o}pke}, G. 2009, \prc, 79, 014002

\bibitem[{{R{\"o}pke}(2011)}]{Ropke:2011tr}
{R{\"o}pke}, G. 2011, Nucl. Phys. A, 867, 66

\bibitem[{{R{\"o}pke}(2015)}]{Ropke:2014fia}
{R{\"o}pke}, G. 2015, Phys. Rev. C, 92, 054001

\bibitem[{{R{\"o}pke}(2020)}]{Ropke:2020peo}
{R{\"o}pke}, G. 2020, Phys. Rev. C, 101, 064310

\bibitem[{{R{\"o}pke} {et~al.}(1982){R{\"o}pke}, {M{\"u}nchow}, \&
  {Schulz}}]{1982NuPhA.379..536R}
{R{\"o}pke}, G., {M{\"u}nchow}, L., \& {Schulz}, H. 1982, \nphysa, 379, 536

\bibitem[{R\"opke {et~al.}(2021)R\"opke, Natowitz, \& Pais}]{Ropke:2020hbm}
R\"opke, G., Natowitz, J.~B., \& Pais, H. 2021, Phys. Rev. C, 103, 061601

\bibitem[{R\"opke {et~al.}(1983)R\"opke, Schmidt, M\"unchow, \&
  Schulz}]{Ropke:1983lbc}
R\"opke, G., Schmidt, M., M\"unchow, L., \& Schulz, H. 1983, Nucl. Phys. A,
  399, 587

\bibitem[{{Royer}(2008)}]{Royer2008a}
{Royer}, G. 2008, \nphysa, 807, 105

\bibitem[{Schmidt {et~al.}(1990)Schmidt, R\"opke, \& Schulz}]{Schmidt:1990oyr}
Schmidt, M., R\"opke, G., \& Schulz, H. 1990, Annals Phys., 202, 57

\bibitem[{{Seitenzahl} {et~al.}(2009){Seitenzahl}, {Townsley}, {Peng}, \&
  {Truran}}]{2009ADNDT..95...96S}
{Seitenzahl}, I.~R., {Townsley}, D.~M., {Peng}, F., \& {Truran}, J.~W. 2009,
  Atomic Data and Nuclear Data Tables, 95, 96

\bibitem[{Siegel(2022)}]{Siegel:2022upa}
Siegel, D.~M. 2022, Nature Rev. Phys., 4, 306

\bibitem[{{Sneden} {et~al.}(1996){Sneden}, {McWilliam}, {Preston}, {Cowan},
  {Burris}, \& {Armosky}}]{sneden1996a}
{Sneden}, C., {McWilliam}, A., {Preston}, G.~W., {et~al.} 1996, \apj, 467, 819

\bibitem[{{Sumiyoshi} \& {R{\"o}pke}(2008)}]{2008PhRvC..77e5804S}
{Sumiyoshi}, K. \& {R{\"o}pke}, G. 2008, \prc, 77, 055804

\bibitem[{{Terasawa} \& {Sato}(1990)}]{1990ApJ...362L..47T}
{Terasawa}, N. \& {Sato}, K. 1990, \apjl, 362, L47

\bibitem[{Typel {et~al.}(2010)Typel, R{\"o}pke, Kl{\"a}hn, Blaschke, \&
  Wolter}]{Typel:2009sy}
Typel, S., R{\"o}pke, G., Kl{\"a}hn, T., Blaschke, D., \& Wolter, H.~H. 2010,
  Phys. Rev. C, 81, 015803

\bibitem[{Wanajo(2013)}]{Wanajo:2013hba}
Wanajo, S. 2013, Astrophys. J. Lett., 770, L22

\bibitem[{Wanajo {et~al.}(2018)Wanajo, Sekiguchi, Nishimura, Kiuchi, Kyutoku,
  \& Shibata}]{Wanajo:2018xex}
Wanajo, S., Sekiguchi, Y., Nishimura, N., {et~al.} 2018, JPS Conf. Proc., 23,
  012033

\bibitem[{Wang {et~al.}(2021)Wang, Huang, Kondev, Audi, \&
  Naimi}]{Wang:2021xhn}
Wang, M., Huang, W.~J., Kondev, F.~G., Audi, G., \& Naimi, S. 2021, Chin. Phys.
  C, 45, 030003

\bibitem[{{Watson} {et~al.}(2019){Watson}, {Hansen}, {Selsing}, {Koch},
  {Malesani}, {Andersen}, {Fynbo}, {Arcones}, {Bauswein}, {Covino}, {Grado},
  {Heintz}, {Hunt}, {Kouveliotou}, {Leloudas}, {Levan}, {Mazzali}, \&
  {Pian}}]{watson2019a}
{Watson}, D., {Hansen}, C.~J., {Selsing}, J., {et~al.} 2019, \nat, 574, 497

\bibitem[{{Xylakis-Dornbusch} {et~al.}(2024){Xylakis-Dornbusch}, {Hansen},
  {Beers}, {Christlieb}, {Ezzeddine}, {Frebel}, {Holmbeck}, {Placco},
  {Roederer}, {Sakari}, \& {Sneden}}]{2024A&A...688A.123X}
{Xylakis-Dornbusch}, T., {Hansen}, T.~T., {Beers}, T.~C., {et~al.} 2024, \aap,
  688, A123

\bibitem[{Yudin {et~al.}(2019)Yudin, Hempel, Blinnikov, Nadyozhin, \&
  Panov}]{Yudin:2018ckx}
Yudin, A., Hempel, M., Blinnikov, S., Nadyozhin, D., \& Panov, I. 2019, Mon.
  Not. Roy. Astron. Soc., 483, 5426

\bibitem[{Zubarev {et~al.}(1996/1997)Zubarev, Morozov, \& R{\"o}pke}]{Zubarev}
Zubarev, D., Morozov, V., \& R{\"o}pke, G. 1996/1997, {Statistical Mechanics of
  Non-equilibrium Processes I/II} (Wiley)

\end{thebibliography}

\appendix
\section{Gross structure of solar abundances}

In Table~\ref{tab.1}, we give values for the accumulated mass fraction $\hat X_{\hat A}$, defined in Eq.~(\ref{XA})
and the $\hat A$-metallicity $M_{\hat A}$, defined in Eq.~(\ref{MA}) as plotted in Figs.~\ref{fig:1} and \ref{fig:XAshelln}. 
Solar mass fractions $\hat X^\odot_{\hat A}$ and  $\hat A$-metallicity $M^\odot_{\hat A}$ are obtained from \citet{2009LanB...4B..712L}. The primordial mass fractions $\hat X^\mathrm{prim}_{\hat A}$ and  $\hat A$-metallicity $M^\mathrm{prim}_{\hat A}$ assume parameter values of $T\,{=}\,5.266\,\mathrm{MeV}$, $\mu_n\,{=}\,940.317\,\mathrm{MeV}$ and $\mu_p=845.069\,\mathrm{MeV}$.

\begin{table}[h]
\caption{Accumulated mass fractions$\hat X_{\hat A}$ and $\hat A$-metallicities $M_{\hat A}$.}  
\label{tab.1}
\centering 
\begin{tabular}{RRRRR}      
\hline\hline\\[-2ex]
\hat A & \log \hat X^\odot_{\hat A}    &  \log M^\odot_{\hat A} & \log \hat X^{\rm prim}_{\hat A}    &  \log M^{\rm prim}_{\hat A} \\[1ex]
\hline\\[-1ex]
0	&-0.1523	&0.0000		& \ldots & 0.0000\\
4	&-0.5555	&-0.5291	& -2.1068& -2.0424\\
8	&-8.2524	&-1.7580	& -3.1644& -2.9030 \\
12	&-2.4070	&-1.7580	& -4.5508& -3.2476\\
16	&-2.1284	&-1.8684	& -3.4172& -3.2698 \\
20	&-2.6028	&-2.2147	& -4.6030 & -3.8108\\
24	&-3.1198	&-2.4432	& -4.0245& -3.8873\\
28	&-3.1011	&-2.5459	& -4.9128& -4.4543\\
32	&-3.4024	&-2.6876	& -4.9323& -4.6400  \\
36	&-3.9822	&-2.7807	& -5.2269& -4.9501 \\
40	&-4.2017	&-2.8089	& -6.0153& -5.2767 \\
44	&-5.6959	&-2.8268	& -5.9716& -5.3642\\
48	&-5.3849	&-2.8274	& -6.3886& -5.4874\\
52	&-3.9617	&-2.8286	& -6.1494& -5.5457\\
56	&-2.8715	&-2.8618	& -6.2884& -5.6701\\
60	&-4.5829	&-4.5167	& -6.1171& -5.7897\\
64	&-5.5298	&-5.3657	& -6.7205& -6.0659\\
68	&-6.2388	&-5.8679	& -7.1405& -6.1747\\
72	&-6.7211	&-6.1088	& -7.3338& -6.2244\\
76	&-7.0309	&-6.2303	& -7.5739& -6.2595\\
80	&-6.8766	&-6.3052	& -7.5629& -6.2811\\
84	&-6.9336	&-6.4408	& -7.6142& -6.3044\\
88	&-7.1233	&-6.6093	& -7.4308& -6.3263 \\
92	&-7.8379	&-6.7681	& -7.2421& -6.3618\\
96	&-8.2692	&-6.8067	& -6.8260 & -6.4231\\
100	&-8.3142	&-6.8219	& -7.4362& -6.6418\\
104	&-8.3692	&-6.8362	& -7.8646& -6.7177\\
108	&-8.4435	&-6.8491	& -8.1783& -6.7499\\
112	&-8.3928	&-6.8603	& -8.3965& -6.7664\\
116	&-8.1629	&-6.8732	& -8.5154& -6.7767\\
120	&-8.2007	&-6.8961	& -8.6020 & -6.7847\\
124	&-7.9386	&-6.9182	& -8.6037& -6.7913\\
128	&-7.6523	&-6.9617	& -8.5831& -6.7981\\
132	&-7.9881	&-7.0607	& -8.4733& -6.8053\\
136	&-7.7132	&-7.1154	& -8.3179& -6.8147\\
140	&-8.1821	&-7.2417	& -8.0678& -6.8285\\
144	&-8.7119	&-7.2946	& -7.7928& -6.8543\\
148	&-9.0105	&-7.3116    & -7.6929& -6.9075 \\
152	&-8.9770    &-7.3203	& -8.2115& -6.9852\\
156	&-8.9190    &-7.3300	& -8.5854& -7.0118\\
160	&-8.7799	&-7.3414	& -8.8581& -7.0236\\
164	&-8.7875    &-7.3575	& -9.0451& -7.0299\\
168	&-9.0449	&-7.3739    & -9.1690& -7.0342\\
172	&-8.9853	&-7.3833	& -9.2471& -7.0374\\
176	&-9.1763	&-7.3943	& -9.2859& -7.0401\\
180	&-9.1656	&-7.4011	& -9.2979& -7.0425\\
\hline
\end{tabular}
\end{table}

\begin{table}[h]
\caption{Accumulated mass fractions$\hat X_{\hat A}$ and $\hat A$-metallicities $M_{\hat A}$.}  
\label{tab.1}
\centering 
\begin{tabular}{RRRRR}      
\hline\hline\\[-2ex]
\hat A & \log \hat X^\odot_{\hat A}    &  \log M^\odot_{\hat A} & \log \hat X^{\rm prim}_{\hat A}    &  \log M^{\rm prim}_{\hat A} \\[1ex]
\hline\\[-1ex]
184	&-9.0851	&-7.4091	& -9.2786& -7.0446\\
188	&-8.4962	&-7.4183	& -9.2322& -7.0475\\
192	&-8.0903	&-7.4562	& -9.1452& -7.0504\\
196	&-8.4125	&-7.5710	& -9.0142& -7.0537\\
200	&-8.7710    &-7.6385	& -8.8327& -7.0586\\
204	&-8.0595	&-7.6718	& -8.6202& -7.0660\\
208	&-7.9157	&-7.9005	& -8.3788& -7.0783\\
212	&\ldots		&-9.3651    & -8.0964& -7.1006\\
216	&\ldots		&-9.3651	& -8.2286& -7.1468\\
220	&\ldots		&-9.3651    & -8.6321& -7.1844\\
224	&\ldots		&-9.3651    & -8.8996& -7.2002\\
228	&\ldots		&-9.3651	& -9.0644& -7.2089\\
232	&-9.5072	&-9.3651	& -9.1504& -7.2150\\
236	&-9.9195	&-9.9195	& -9.1806& -7.2201\\
240 &\ldots		&\ldots		& -9.1699& -7.2249\\
244 & \ldots & \ldots &-9.1305& -7.2299 \\ 
248 & \ldots & \ldots & -9.0676& -7.2353 \\ 
252 & \ldots & \ldots & -8.9829& -7.2418 \\ 
256 & \ldots & \ldots & -8.8717& -7.2497 \\ 
260 & \ldots & \ldots & -8.7252& -7.2602 \\ 
264 & \ldots & \ldots & -8.5343& -7.2754 \\ 
268 & \ldots & \ldots & -8.7735& -7.2999 \\ 
272 & \ldots & \ldots & -9.0477& -7.3148 \\ 
276 & \ldots & \ldots & -9.2409& -7.3229 \\ 
280 & \ldots & \ldots & -9.3691& -7.3282 \\
284 & \ldots & \ldots & -9.4461& -7.3322 \\ 
288 & \ldots & \ldots & -9.4848& -7.3356 \\ 
292 & \ldots & \ldots & -9.4959& -7.3387 \\ 
296 & \ldots & \ldots & -9.4889& -7.3417 \\ 
300 & \ldots & \ldots & -9.4711& -7.3448 \\ 
304 & \ldots & \ldots & -9.4482& -7.3481 \\ 
308 & \ldots & \ldots & -9.4243& -7.3515 \\ 
312 & \ldots & \ldots & -9.4017& -7.3552 \\ 
316 & \ldots & \ldots & -9.3806& -7.3591 \\
320 & \ldots & \ldots & -9.3598& -7.3633 \\ 
324 & \ldots & \ldots & -9.3358& -7.3677 \\
328 & \ldots & \ldots & -9.3038& -7.3724 \\ 
332 & \ldots & \ldots & -9.2566& -7.3775 \\
336 & \ldots & \ldots & -9.1856& -7.3833 \\ 
340 & \ldots & \ldots & -9.0802& -7.3902 \\
344 & \ldots & \ldots & -8.9283& -7.3991 \\ 
348 & \ldots & \ldots & -8.7156& -7.4122 \\
352 & \ldots & \ldots & -8.4272& -7.4343 \\ 
356 & \ldots & \ldots & -8.0566& -7.4809 \\
360 & \ldots & \ldots & -8.2879& -7.6149 \\ 
364 & \ldots & \ldots & -8.7160& -7.7186 \\
368 & \ldots & \ldots & -9.0654& -7.7646 \\ 
372 & \ldots & \ldots & -9.3259& -7.7869 \\
376 & \ldots & \ldots & -9.4967& -7.7997 \\ 
380 & \ldots & \ldots & -9.5922& -7.8085 \\
384 & \ldots & \ldots & -9.6293& -7.8157 \\ 
388 & \ldots & \ldots & -9.6242& -7.8224 \\
392 & \ldots & \ldots & -9.5897& -7.8293 \\ 
396 & \ldots & \ldots & -9.5365& -7.8369 \\
400 & \ldots & \ldots & -9.4723& -7.8457 \\
\hline
\end{tabular}
\end{table}

\end{document}